\documentclass[prd,aps,preprint,showpacs,preprintnumbers,amsmath,amssymb,nofootinbib]{revtex4}
\usepackage{graphicx}
\usepackage{subfig}
\usepackage{bm}
\usepackage{ulem}

\def\wisk#1{\ifmmode{#1}\else{$#1$}\fi}
\def\lsim{\wisk{_<\atop^{\sim}}}

\def\1{\'\i}

\begin{document}

\title{The effect of Dark Matter and Dark Energy interactions on the
 peculiar velocity field and the kinetic Sunyaev-Zel'dovich effect}

\author{Xiao-Dong Xu, Bin Wang,  Pengjie Zhang}

\affiliation{Department of Physics and Astronomy,
Shanghai Jiao Tong University, Shanghai 200240,
China}

\author{ Fernando Atrio-Barandela}

\affiliation{Departamento de F\1sica Fundamental,
Universidad de Salamanca, Spain}

\begin{abstract}

The interaction between Dark Matter and Dark Energy has been proposed
as a mechanism to alleviate the coincidence problem. We analyze its
effects on the evolution of the gravitational and the peculiar velocity
fields. We find that for different model parameters peculiar velocities
vary from a factor five times smaller to two times larger than in the
$\Lambda$CDM cosmological model at the same scales. We propose two new
observables sensitive to such interactions based on their
effect on the velocity field. We compare the effects on
peculiar velocities with those on the Integrated Sachs-Wolfe effect
demonstrating that velocities are more sensitive to the interaction.
We show that the current upper limits on the amplitude of
the kinetic Sunyaev-Zel'dovich power spectrum of temperature
anisotropies provide constraints on the coupling within the dark
sectors that are consistent with those obtained previously from
the Cosmic Microwave Background and galaxy clusters. In particular,
we show that Atacama Cosmology Telescope and South Pole Telescope data
favor the decay of Dark Energy into Dark Matter, as required
to solve the coincidence problem.
\end{abstract}


\maketitle

\section{Introduction}

Luminosity distances derived from Type-Ia supernovae (SNIa) were
the first cosmological observations to establish that today the
expansion of the Universe is being accelerated \cite{Perlmutter1998,
Riess1998, Perlmutter1999, Tonry2003, Riess2004, Astier2006, Riess2007},
driven by a so-called dark energy (DE) with equation of state (EoS)
parameter $w\approx -1$. Observations of the temperature anisotropies
of the Cosmic Microwave Background (CMB) indicated that the dominant
energy component, with $\sim 70$\% of the total energy density is DE,
$\sim 26$\% cold dark matter (DM) and a small fraction (4\%) of
baryonic matter \cite{Hinshaw2012, PlanckCollaboration2013b}. The nature of
both DM and DE are unknown. They couple to baryons and radiation only
through gravity but in the context of field theory other interactions
within the dark sector can also exist \cite{Micheletti2009, Micheletti2009a}.
A DM-DE interaction provides a natural way to alleviate the coincidence
problem, which embarrasses the standard $\Lambda$CDM cosmology
\cite{Amendola2000, Amendola2003, Amendola2006, Pavon2005, Campo2008,
Boehmer2008, Chen2008}. Also, the appropriate interaction can accommodate
an effective DE EoS with $w<-1$ at the present time \cite{Wang2005, Das2006}.
A non-gravitational coupling in the dark sector will affect significantly the
expansion history of the Universe and the evolution of density perturbations,
changing their growth. The possibility of the DE-DM interaction has been
widely discussed in the literature \cite{Amendola2000, Amendola2003,
Amendola2006, Pavon2005, Campo2008, Olivares2005, Olivares2008a,
Boehmer2008, Chen2008, Valiviita2008, He2008a, Corasaniti2008,
Jackson2009, Pavon2007, Wang2007, Wang2007a, Simpson2010, Zimdahl2005,
Guo2007, Feng2008, Valiviita2009, Xia2009, He2009, Martinelli2010,
Honorez2010a, He2008, He2009a, Caldera-Cabral2009, He2010, Bertolami2007,
Bertolami2009, Abdalla2007, Abdalla2009}. Thus, determining the
existance of DM-DE interactions is an observational endeavour that
could provide an interesting insight into the nature of the dark sector.

Several authors have looked for observational signatures of the
interaction using Wilkinson Microwave Anisotropy Probe ({\it WMAP}),
SNIa and Baryon Oscillation Observation (BAO) data \cite{Simpson2010,
Zimdahl2005, Guo2007, Feng2008, Valiviita2009, Xia2009, He2009,
Martinelli2010} together with complementary probes on the growth of
cosmic structures \cite{Olivares2006, He2009a, Caldera-Cabral2009,
He2010, Bertolami2007, Bertolami2009, Abdalla2007, Abdalla2009}.
Until now the data have provided only upper limits on the amplitude
of the interaction, requiring the strength of the coupling to be
$\le 10^{-3}$ \cite{Abdalla2007, He2010a, Xu2011, Salvatelli2013}.
Therefore, it is necessary to search for more sensitive probes.
For instance, the interaction changes the time evolution
of the gravitational potential \cite{Olivares2006,He2009} and leaves
a signature in the late Integrated Sachs-Wolfe (ISW) effect
\cite{Olivares2008b, He2009, He2010a}. Due to cosmic variance,
the statistical power of the late-time ISW to constrain DM-DE
interactions is limited. Nevertheless, since the coupling between
the dark sectors also influences the absolute value of the gravitational
potential and changes the peculiar motion of the different matter particles,
it provides two new observational tests. First, matter peculiar
velocities can be measured using galaxy surveys and
kinetic Sunyaev-Zel'dovich (kSZ) temperature anisotropies.
Depending on model parameters, the interaction changes the
amplitude of matter peculiar velocities by a factor two to five with
respect to the fiducial $\Lambda$CDM model. While these variations
are significant, they are still within the upper limit set by
{\it Planck} \cite{PlanckCollaboration2013a}. Second,
the diffuse intergalactic medium and the unresolved cluster
population generate temperature anisotropies on the CMB at
small scale via the conventional kSZ effect and any
change of the baryon peculiar velocity field
will also change the power spectrum of the kSZ anisotropies.
In this paper we will show that both observational tests
are potentially powerful probes of DM-DE interactions.

In general, the kSZ effect is sensitive to the dynamical
evolution of matter in the Universe. Large scale bulk flows and
the mean pairwise velocity dispersion of clusters have already
been reported \cite{Kashlinsky2008, Kashlinsky2010, Kashlinsky2011,
Hand2012, Mroczkowski2012}. The kSZ effect is also a potential
probe of reionization, the radial inhomogeneities in the
Lema{\^\i}tre-Tolman-Bondi \cite{Bull2011}, the missing baryon
problem \cite{Genova-Santos2009}, the dark flow \cite{Zhang2010}, etc.
Throughout the paper we will consider only a flat cosmological
model with the following cosmological parameters:
Hubble constant $H_0=67.11 \mathrm{km~s^{-1}~Mpc^{-1}}$,
baryon abundance $\Omega_bh^2=0.0221$, DE fraction
$\Omega_{\mathrm{d}}=0.68$ (for $\Lambda$CDM model,
$\Omega_{\Lambda}=0.68$), scalar spectral index $n_s=0.9624$
and amplitude of primordial curvature perturbation $10^9A_s=2.215$.
Briefly, in Sec.~2 we will introduce the equations describing the
evolution of DM and DE density perturbations. In Sec.~3, we will
study the effect of the interaction on the evolution of the
gravitational field and its contribution to the Integrated
Sachs Wolfe effect. In Sec.~4, we will analyze the effect on
peculiar velocities and in Sec.~5 we will compute the power spectrum
of the CMB temperature anisotropies generated by the kSZ effect. Finally,
in Sec.~6 we will summarize our main results and present our conclusions.

\section{The model of the interaction between DE and DM}

The formalism describing the evolution of matter and DE density
perturbations without \cite{Kodama1984, Ma1995} and with DM-DE
interaction \cite{He2009a} is well established. If DM and DE are
coupled with each other, the energy-momentum tensor
$T^{\mu\nu}_{\lambda}$ of each individual
component $\lambda=(\mathrm{c,d})$ is no longer conserved.
Instead,
\begin{equation}
\nabla_{\mu}T^{\mu\nu}_{\lambda} = Q^{\nu}_{\lambda},
\label{emTensor}
\end{equation}
where $Q^{\nu}_{\lambda}$ is the four vector governing the
energy-momentum transfer between dark components and the subscripts
$\mathrm{c,d}$ refer to DM and DE, respectively.
Since DM and DE couple to all other energy densities in the Universe only
through gravity, the energy-momentum tensor of DM+DE is conserved, i.e.,
$\sum_{\lambda}\nabla_{\mu}T^{\mu\nu}_{\lambda}=0$.
Then, $Q^{\nu}_{\mathrm{c}}=-Q^{\nu}_{\mathrm{d}}$.

Assuming spatially flat Friedmann-Robertson-Walker background,
from eq~(\ref{emTensor}) we can derive the equations of evolution
of the mean DM and DE densities
\begin{equation}
\dot{\rho_{\mathrm{c}}}+3H\rho_{\mathrm{c}}=Q_{\mathrm{c}}, \quad
\dot{\rho_{\mathrm{d}}}+3H(1+w)\rho_{\mathrm{d}}=Q_{\mathrm{d}},
\end{equation}
where $Q_{\lambda}=aQ_{\lambda}^0$ indicates the energy transfer,
$Q_{\lambda}^0$ is the time component of the four vector
$Q^{\nu}_{\lambda}$ and $a$ the scale factor. Dots
represent derivatives with respect to the coordinate time.
Since the physical properties of DM and DE at the
present moment are unknown, we cannot derive the
precise form of the interaction from first
principles. For simplicity, we will consider a
phenomenological description and thus write
the interaction between DE and DM as a linear
combination of energy densities of the dark sectors
$Q_{\mathrm{c}}=3H(\xi_1\rho_{\mathrm{c}}+
\xi_2\rho_{\mathrm{d}})$ \cite{He2008a, He2009a}.
With this functional form, $\xi_1$ and
$\xi_2$ are dimensionless parameters. To simplify the problem,
we will assume that these coefficients are independent of time
and the EoS parameter $w$ is constant.

\begin{figure}[tbp]
\subfloat[]{
    \includegraphics[width=0.45\textwidth]{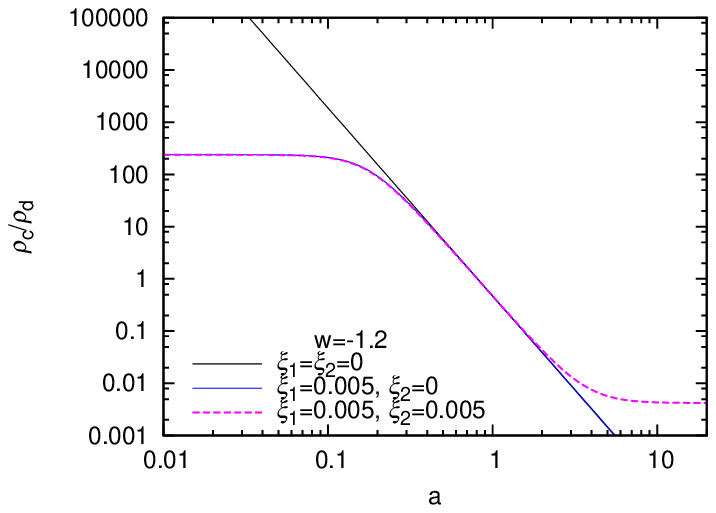}
}
\subfloat[]{
    \includegraphics[width=0.45\textwidth]{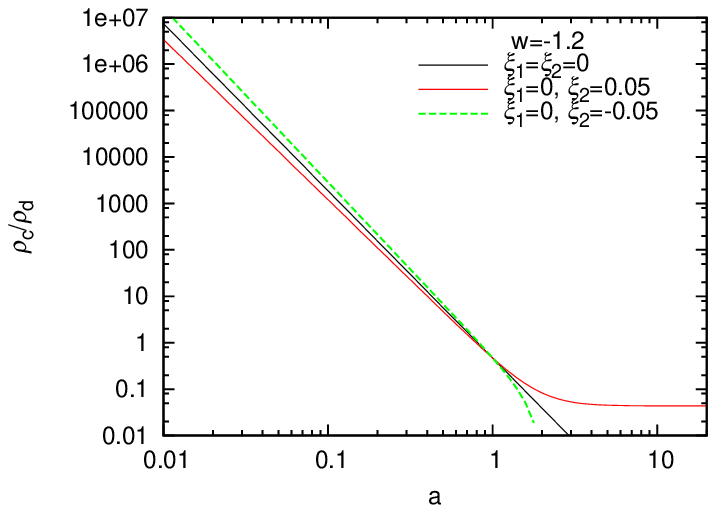}
}
\caption{The evolution of the DM-DE energy density ratio
$r \equiv \rho_{\mathrm{c}}/\rho_{\mathrm{d}}$ in a model
with an interaction $Q_c=3H(\xi_1\rho_c+\xi_2\rho_d)$,
for different coupling constants.}
\label{ratio}
\end{figure}

If the interaction parameters are positive, the DE transfers
energy to the DM while if they are negative, the transfer is
in the opposite direction. While the sign of the interaction is
unknown, thermodynamic considerations based on the second
principle require the coupling constants to be positive
\cite{Pavon2007}. Also, to avoid the unphysical solution of
a negative dark energy density ($\rho_d<0$) in the early Universe
when $Q_c=3H\xi_1\rho_c$ or $Q_c=3H\xi(\rho_c+\rho_d)$
the couplings $\xi_1,\xi$ must be positive \cite{He2008}.
In Fig.~\ref{ratio} we represent
the ratio of the mean energy densities $r \equiv \rho_d/\rho_c$.
In Fig.~\ref{ratio}a we represent the ratio with $\xi_2=0$
and $\xi_1=\xi_2$; in Fig.~\ref{ratio}b we show $\xi_1=0$.
There are two attractor solutions with $r=const$, in the past,
if $\xi_1>0$, and in the future, if $\xi_2>0$
\cite{Amendola2003,Pavon2005,Olivares2005}.
Compared with the $\Lambda$CDM model,  positive couplings
gives a slower evolution $r$, i.e., there exists a long period in
the evolution of the Universe where the DM and DE densities
are comparable, alleviating the coincidence problem.

We can further reduce the parameter space if we take into
account that, at the background level and at first order in
perturbation theory, models with ($\xi_1=\xi_2 \neq 0$)
and ($\xi_1 \neq 0$, $\xi_2=0$) show similar behavior.
Also, since the fraction of DE at early times is
small compared to the DM, models with interaction kernels
$\xi_1 \neq \xi_2$ behave very similarly to models with $\xi_2=0$.
If $Q_c=3H\xi_1\rho_c$ or
$Q_c=3H\xi(\rho_c+\rho_d)$, curvature perturbations diverge
if $w > -1$ so we restrict
our study to $w < -1$. A discussion on the origin of this instability
can be found in \cite{He2008a, Xu2011}. If $Q_c=3H\xi_2\rho_d$,
there are no instabilities in the matter density perturbations, so we
can consider EoS parameter with values larger and smaller than
$w=-1$.  To summarize, we will study two interaction kernels:
$Q_c=3H\xi_1\rho_c$ and $Q_c=3H\xi_2\rho_d$; in the first
case, $\xi_1>0$ and $w<-1$; in the second, $\xi_2$ can have
positive and negative value and $w$ can be smaller or larger than -1.
Other expressions, like the product of DM and DE densities,
have also been discussed in the literature, but since they
require a different treatment of first order perturbations
we shall not consider them here.

The general gauge-invariant perturbation equations for DM
and DE are \cite{He2010a}
\begin{align}
D_{\mathrm{c}}' =& -kU_{\mathrm{c}} + 3\mathcal{H}(\xi_1+\xi_2/r)\Psi -
3(\xi_1+\xi_2/r)\Phi' + 3\mathcal{H}\xi_2(D_{\mathrm{d}}-D_{\mathrm{c}})/r,
\label{eq:perturbations1}\\
U_{\mathrm{c}}'\;\; =& -\mathcal{H}U_{\mathrm{c}} + k\Psi - 3\mathcal{H}(\xi_1+\xi_2/r)U_{\mathrm{c}},
\label{eq:perturbations2}\\
\begin{split}
D_{\mathrm{d}}' =& -3\mathcal{H}(C_{\mathrm{e}}^2-w)D_{\mathrm{d}} +
\{3w' + 9\mathcal{H}(C_{\mathrm{e}}^2-w)(\xi_1r+\xi_2+1+w)\}\Phi\\
&- 9\mathcal{H}^2(C_{\mathrm{e}}^2-C_{\mathrm{a}}^2)\frac{U_{\mathrm{d}}}{k} +
3(\xi_1r+\xi_2)\Phi' - 3\mathcal{H}(\xi_1r+\xi_2)\Psi\\
&+ 3\mathcal{H}\xi_1r(D_{\mathrm{d}}-D_{\mathrm{c}}) -
9\mathcal{H}^2(C_{\mathrm{e}}^2-C_{\mathrm{a}}^2)(\xi_1r+\xi_2)\frac{U_{\mathrm{d}}}{(1+w)k} - kU_{\mathrm{d}},
\label{eq:perturbations3}\end{split}\\
\begin{split}
U_{\mathrm{d}}' =& -\mathcal{H}(1-3w)U_{\mathrm{d}} - 3kC_{\mathrm{e}}^2(\xi_1r+\xi_2+1+w)\Phi
+ 3\mathcal{H}(C_{\mathrm{e}}^2-C_{\mathrm{a}}^2)(\xi_1r+\xi_2)\frac{U_{\mathrm{d}}}{1+w}\\
&+ 3\mathcal{H}(C_{\mathrm{e}}^2-C_{\mathrm{a}}^2)U_{\mathrm{d}} + kC_{\mathrm{e}}^2D_{\mathrm{d}} + (1+w)k\Psi +
3\mathcal{H}(\xi_1r+\xi_2)U_{\mathrm{d}}.
\label{eq:perturbations4}\end{split}
\end{align}
Primes denote derivation with respect to the
conformal time, $\mathcal{H}\equiv a'/a$ and
$U_{\lambda} \equiv (1+w_{\lambda})V_{\lambda}$.
$D_{\lambda} \equiv \delta_{\lambda}-
\frac{\rho'_{\lambda}}{\rho_{\lambda}\mathcal{H}}\Phi$
is the gauge invariant density contrast and $C_{\mathrm{e}}^2$,
$C_{\mathrm{a}}^2$ are the effective and adiabatic sound speed,
respectively. In the Newtonian gauge, $\Psi$, $\Phi$ coincide
with the gravitational potential $\psi$, $\phi$ and $V_{\lambda}$
with the peculiar velocity $v_{\lambda}$ \cite{Kodama1984}.
Eqs.~(\ref{eq:perturbations1}-\ref{eq:perturbations4})
coupled with those describing the evolution of the
other energy density perturbations \cite{Ma1995}, allow us
to compute the evolution of the matter
gravitational potentials and the peculiar velocity
of baryons, and to determine the effect of the
DM-DE interaction on the CMB power spectrum.

\begin{figure}[tb]
\subfloat[]{
    \includegraphics[width=0.45\textwidth]{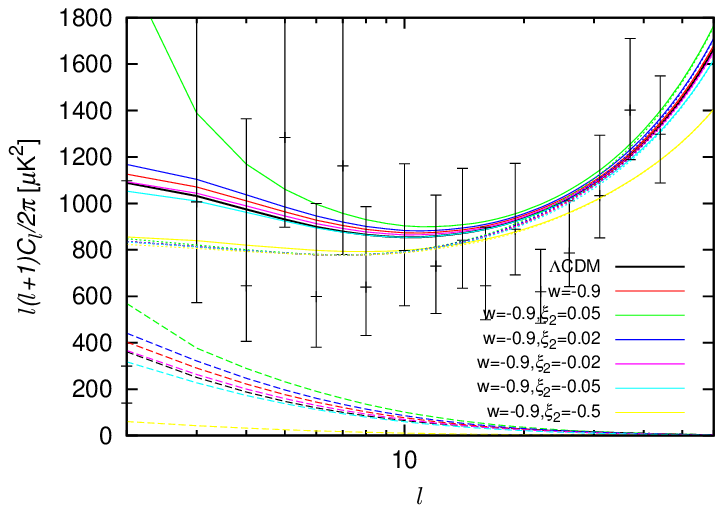}
    \label{wp1x2_isw}
}
\subfloat[]{
    \includegraphics[width=0.45\textwidth]{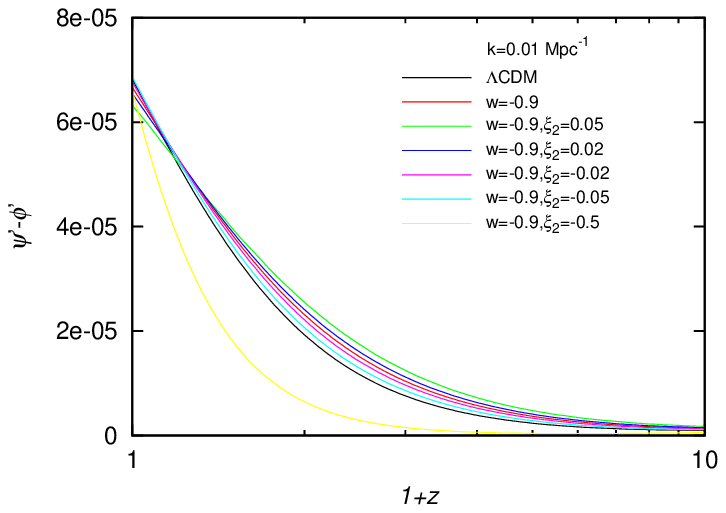}
    \label{wp1x2_phidot}
} \caption{ (a) Solid lines represent the
radiation power spectrum of the CMB at large
scales for the $\Lambda$CDM model and the
phenomenological DM-DE interacting model with
different couplings; the data and error bars were
taken from {\it Planck} 2013 first data release
\cite{PlanckCollaboration2013f}. Dashed lines
display late ISW effect component of the same
models. Finally, the dotted lines represent the
early ISW effect and SW effect of the same models
which are degenerated for the coupling
proportional to the energy density of DE. (b)
Time derivative of the gravitational potential at
lower redshifts, corresponding to a perturbation
of wavenumber $k=0.01$Mpc$^{-1}$. }
\label{wp1x2}\end{figure}

\begin{figure}[tb]
\subfloat[]{
   \includegraphics[width=0.45\textwidth]{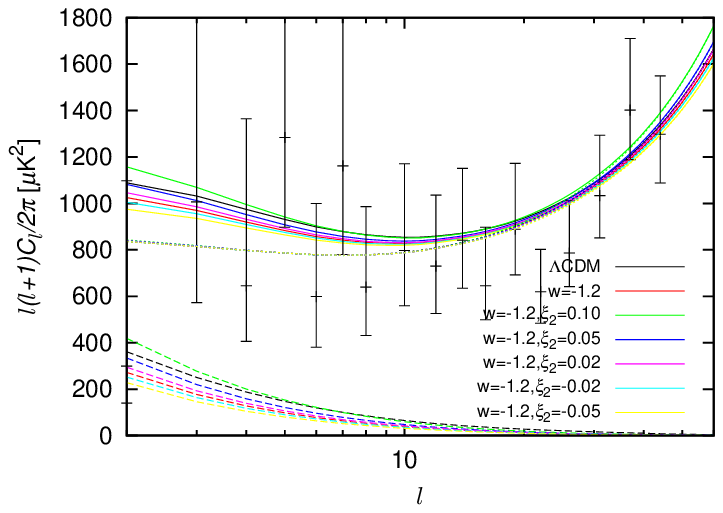}
   \label{wn1x2_isw}
}
\subfloat[]{
   \includegraphics[width=0.45\textwidth]{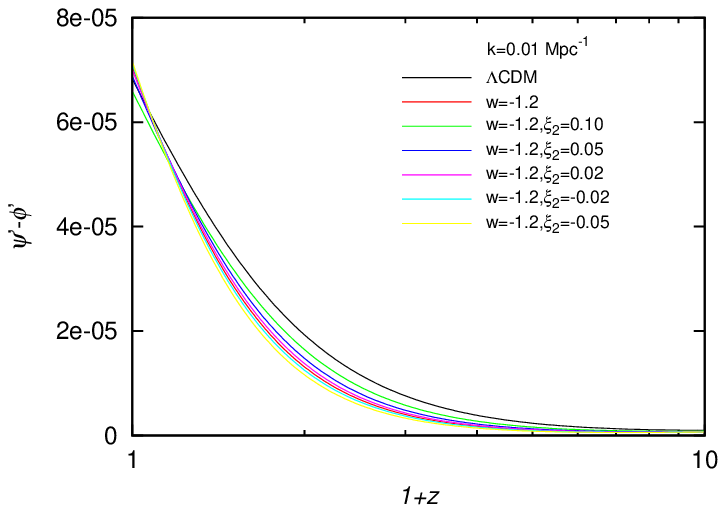}
   \label{wn1x2_phidot}
}
\caption{
Same as Fig.~\ref{wp1x2} but for $w<-1$.
}
\label{wn1x2}\end{figure}

\begin{figure}[tb]
\subfloat[]{
    \includegraphics[width=0.45\textwidth]{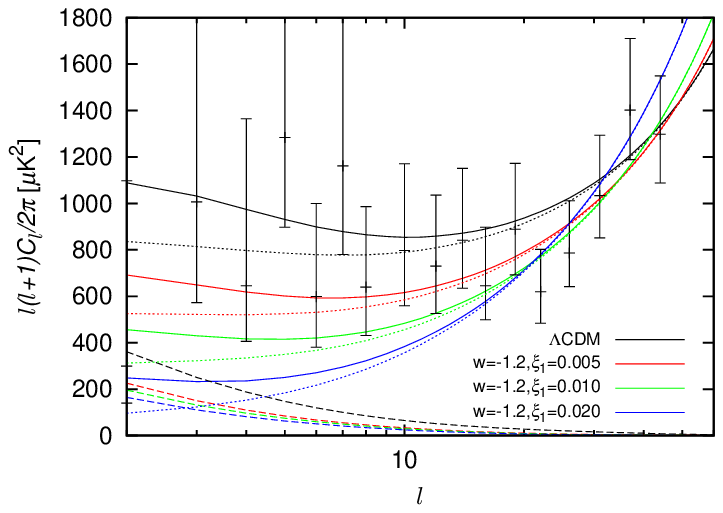}
    \label{wn1x1_sw}
}
\subfloat[]{
    \includegraphics[width=0.45\textwidth]{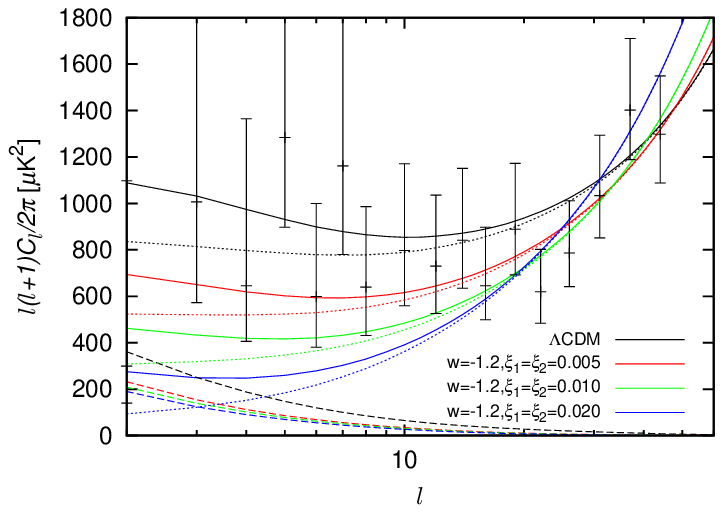}
    \label{wn1xt_sw}
}
\caption{Radiation power spectra (solid lines), late ISW
contribution (dashed lines) and SW+early ISW contributions (dotted
lines). In (a) $\xi_1>0$ and $\xi_2=0$ and in (b) $\xi_1=\xi_2=\xi>0$.}
\label{wn1x1}
\end{figure}

\section{The Integrated Sachs-Wolfe Effect as a probe of DM-DE interactions}

Let us consider the effect of the interaction in the late-time ISW
effect. As indicated in the introduction, the interaction changes
the evolution of the gravitational field, which will result
in differences in the growth of density perturbations compared with
the concordance cosmological model. The different time evolution of the
gravitational potential generates different ISW contributions to
the CMB temperature anisotropies and, if measured, it could
be used to constrain the interaction.

The ISW effect \cite{Sachs1967} is the source of temperature
anisotropies at early and late times \cite{Hu1995}. The latter can
be separated from other CMB anisotropies by cross-correlating with
templates of projected galaxy density \cite{Crittenden1996} and,
therefore, we shall focus on the late-time ISW effect here. The
ISW effect can be simply expressed in terms of integration of
time-evolution of gravitational potential along the line-of-sight
\cite{Sachs1967},
\begin{equation}
\Delta_\ell^{ISW} = \int_{\tau_i}^{\tau_0}
\mathrm{d}\tau j_\ell(k[\tau_0-\tau]) e^{-\kappa(\tau)} (\Psi'-\Phi'),
\label{eq:isw}
\end{equation}
where $j_\ell$ is the spherical Bessel function
and $\kappa$ denotes the optical depth  due to
Thompson scattering. The time evolution of the
gravitational potentials can be described using
Einstein's equations
\begin{align}
\Psi' - \Phi' &= 2\mathcal{H} [ \Phi + \mathcal{T} ]
+ 8\pi Ga^2\sum_i U_i\rho_i/k - \mathcal{T}',\\
\Phi' &= -\mathcal{H}\Phi - \mathcal{HT} - 4\pi Ga^2\sum_i U_i\rho_i/k
\end{align}
where
\begin{align}
\mathcal{T} = \frac{8\pi Ga^2}{k^2} (p_{\gamma}\Pi_{\gamma}
+ p_{\nu}\Pi_{\nu}) .
\end{align}
In this expression $\Pi$ is the anisotropic stress of the
relativistic fluid components, that is negligible after
matter domination. In this limit, $\Phi=-\Psi$. The photon
frequency is shifted on the path to the observer due to time
varying gravitational potentials. At low redshifts, the
contribution is largest at large angular scales.

We solve the evolution equations numerically and compute
the power spectrum of CMB temperature anisotropies
using a version of the Boltzmann code \texttt{CMBEASY}
\cite{Doran2005} publically available, modified according to our model.
In Figs.~\ref{wp1x2}, \ref{wn1x2}, \ref{wn1x1}, we plot
the total radiation power spectrum (solid lines),
the ISW component (dashed lines), the SW+early ISW contribution
(dotted lines) and the time evolution of the gravitational potentials.
In Figs.~\ref{wp1x2},\ref{wn1x2} the interaction is proportional
to the DE ($\xi_1=0$) and the EoS corresponds to quintessence ($w>-1$)
and phantom ($w<-1$), respectively. In Fig.~\ref{wn1x1_sw} the interaction
is proportional to the DM ($\xi_2=0$) and $w<-1$ since only in
this case the growth of matter and dark energy
density perturbations does not diverge. In Fig.~\ref{wn1xt_sw} we
show the interaction proportional to the energy density of the
dark sector with $\xi_1=\xi_2=\xi$ and $w<-1$.

In Fig.~\ref{wp1x2} we have $\xi_2>0$ and the energy is transferred
from DE to DM. A positive coupling gives larger fraction of DE
in the past than in the concordance model,
so that the gravitational potential evolves faster,
giving rise to the enhancement of the ISW effect.
When the coupling is negative,
$\xi_2<0$, we have the opposite behavior: the
fraction of DE is smaller in the past, slowing the
evolution of the gravitational potential and
suppressing the ISW effect. In Fig.~\ref{wn1x2_isw} and
Fig.~\ref{wn1x2_phidot}, with parameters
$\xi_1=0,\;\xi_2>0$ and $w<-1$, we find similar behavior.
Comparing Fig.~\ref{wp1x2_isw} and
Fig.~\ref{wn1x2_isw} for the same value of
$\xi_2$, the ISW is smaller when $w<-1$ than
when $w>-1$. In Fig.~\ref{wn1x1_sw} we plot the results
for $\xi_2=0,\; w<-1$. In this case, increasing the coupling
decreases the ISW effect. Fig.~\ref{wn1xt_sw} shows the ISW
effect in models where the interaction is proportional to the
total energy density of the dark sector, which exhibits similar
properties than the models of Fig.~\ref{wn1x1_sw}.
Comparison of Fig.~\ref{wn1x1_sw} and Fig.~\ref{wn1xt_sw} shows
that the effect of the DE
is small compared to that of the DM. As indicated in Sec.~2,
since at early times the DM density is much larger than that
of the DE, including the DE density in the interaction kernel
does not change the evolution significantly. In general,
when the fraction of DE in the past increases, the gravitational
potential evolves faster, giving rise to a larger ISW effect.
This is similar to what happens in the $\Lambda$CDM model, when
the fraction of $\Omega_\Lambda$ increases.

Together with the late-time ISW effects, there
are early ISW anisotropies, generated since the
redshift of matter-radiation equality until well
after recombination \cite{Hu1995} and the
Sachs-Wolfe(SW) effect generated on the last
scattering surface at large scales. When the
interaction is proportional to the DE energy
density, the early ISW and the SW effect is
little affected since the fraction of DE at high
redshifts is small. This can be seen in
Fig.~\ref{wp1x2_isw} and \ref{wn1x2_isw}. The
dotted lines include the SW and early ISW effects are degenerated and very insensitive to an
interaction that is proportional to the DE
density. When the interaction is proportional to
the DM density, the effect of the interaction in
the early ISW and SW anisotropies is larger since
the DM is the dominant energy density component
after matter-radiation equality. This effect can
be seen analyzing the dotted lines in
Figs.~\ref{wn1x1_sw} and~\ref{wn1xt_sw}. These
figures show that the combined SW and early ISW
power spectra is more sensitive to the
interaction than the late time ISW effect,
represented by dashed lines.

To summarize, the results presented in
Figs.~\ref{wp1x2},\ref{wn1x2},\ref{wn1x1}
indicate that the behavior of the ISW anisotropies
is very sensitive to small changes in the strength of the interaction.
The SW and early ISW effect components of the CMB
temperature anisotropies can not be separated from
other contributions. We can probe the effect of the interaction
by analyzing how the full CMB power spectrum changes and comparing
it with observations. However,
cross-correlating CMB maps with templates of projected density
of galaxies can separate the contribution of the time variation
of the potentials to the late-time ISW effect \cite{Crittenden1996}.
Unfortunately,
due to cosmic variance the statistical power of the ISW effect
to constrain the DM-DE interaction is weak \cite{Olivares2008b}.
Nevertheless, the interaction does not only change the time
evolution of the gravitational potentials but it also
changes the amplitude of potentials \cite{Olivares2006}.
Since the gravitational potential affects the peculiar
motions of all particles the interaction, in turn, will
modulate peculiar velocities of baryons
and their effect on the CMB temperature anisotropies and
both effects can be measured as we will see in the
next two sections.

\section{Probing the interaction with Peculiar Velocities.}

\begin{figure}[tb]
\subfloat[]{
    \includegraphics[width=0.45\textwidth]{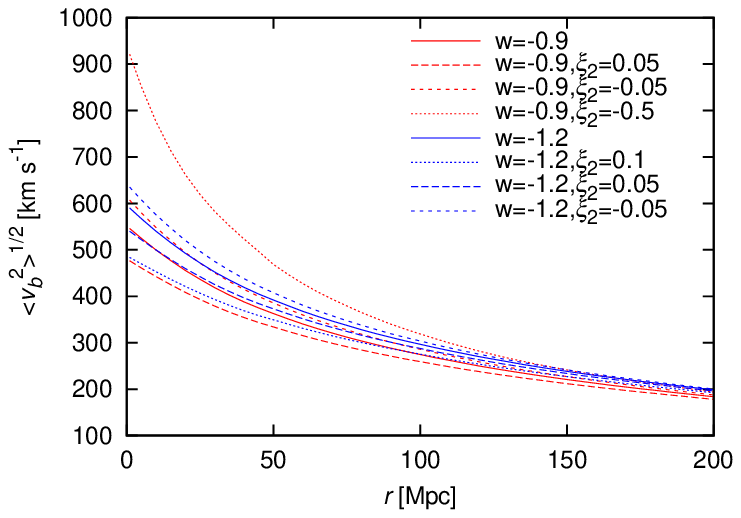}\label{v_x2}
}
\subfloat[]{
    \includegraphics[width=0.45\textwidth]{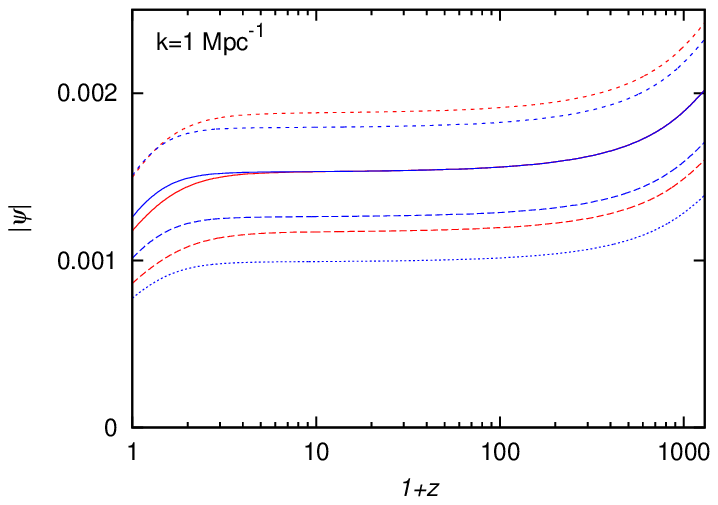}\label{phi_x2}
} \caption{ (a) Root mean square baryon velocity dispersion
averaged on a sphere of radius $r$ \textit{vs.} the size of
the sphere, for an interaction kernel
proportional to DE density. (b) Time evolution of
the gravitational potential for a wavenumber
$k=1$Mpc$^{-1}$. The curves in both panels follow
the same convention.} \label{x2}
\end{figure}

\begin{figure}[tb]
\subfloat[]{
    \includegraphics[width=0.45\textwidth]{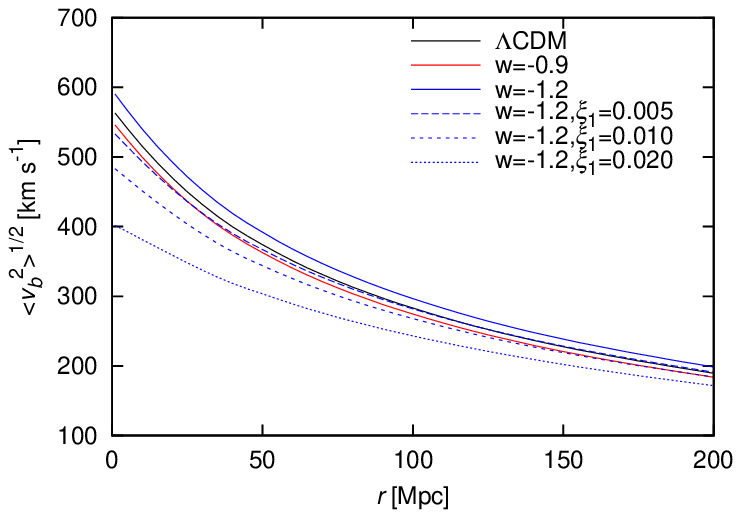}\label{v_x1}
}
\subfloat[]{
    \includegraphics[width=0.45\textwidth]{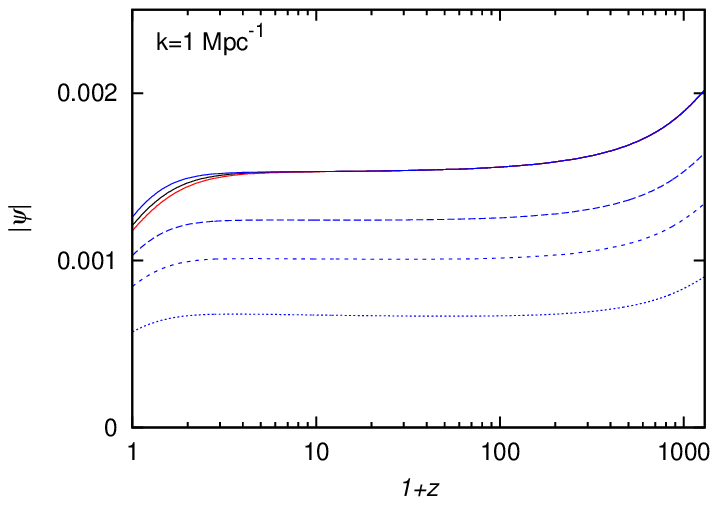}\label{phi_x1}
}\\
\subfloat[]{
    \includegraphics[width=0.45\textwidth]{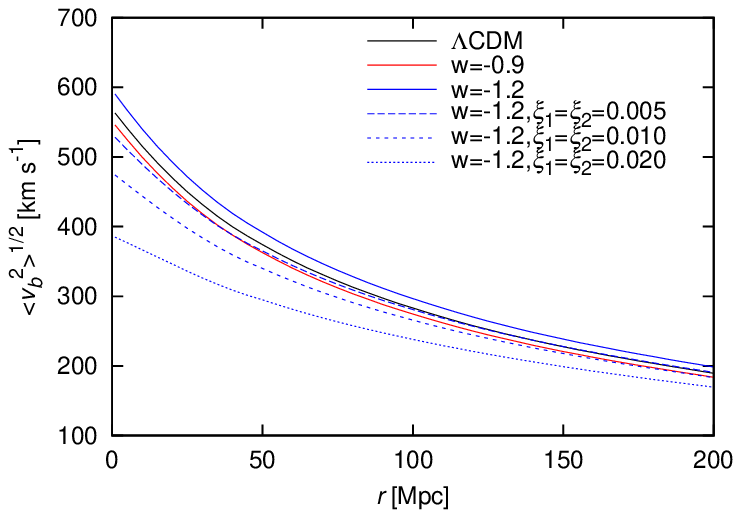}\label{v_xt}
}
\subfloat[]{
    \includegraphics[width=0.45\textwidth]{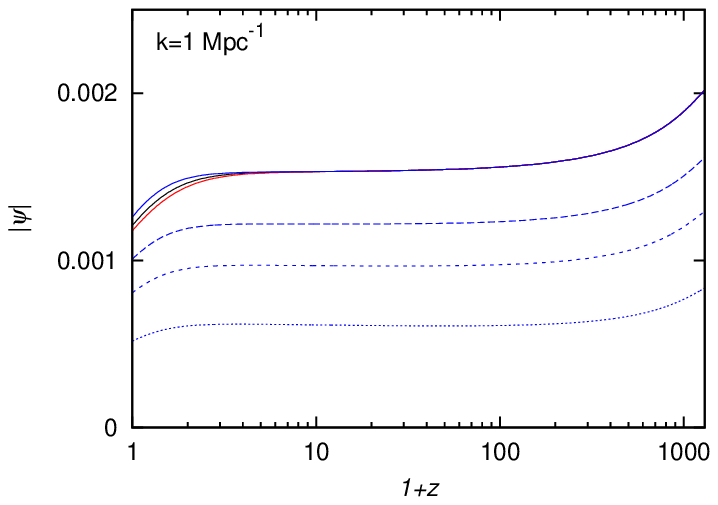}\label{phi_xt}
}
\caption{
Same as in Fig.~\ref{x2} but for: (a) and (b) interaction
kernel proportional to the DM; (c) and (d) interaction kernel
proportional to the total dark sector with $\xi_1=\xi_2$. Again,
only positive values of $\xi_1$ and $\xi$ are considered.}
\label{x1}
\end{figure}

In the standard $\Lambda$CDM cosmology, the
difference between the velocities of baryon and
DM can be neglected at low redshift since baryons trace DM to
high accuracy in the linear regime. In the
presence of DM-DE interactions, baryons do not
trace the DM distribution as perfectly due to the
inertial drag effect of DE on DM. The peculiar
velocity of baryons is also affected through the
influence of the interaction on the curvature
perturbation. Thus, it is possible to extract information of the
coupling from velocity measurements.

In linear perturbation theory, the evolution equation for the velocity of
baryons in Fourier space reads \cite{He2010a}:
\begin{equation}
v_{\mathrm{b}}' = -\mathcal{H}v_{\mathrm{b}} + k\Psi,
\label{eq:vb}
\end{equation}
where the subscript $\mathrm{b}$ refers to baryon.
This equation shows that baryons are accelerated
when falling into the gravitational potential
wells. Although the interaction does not appear explicitly
in Eq.~(\ref{eq:vb}), it affects the evolution of baryon
peculiar velocities through the gravitational
potential $\Psi$. Using Einstein's equations, we obtain \cite{He2008a}
\begin{equation}
\Psi = -\Phi = -\frac{4\pi Ga^2 \sum_i
\rho_i(D_g^i-\rho_i'U_i/(1+w_i)\rho_ik)} {k^2 -
4\pi Ga^2\Sigma_i\rho_i'/\mathcal{H}}.
\end{equation}
The sum is over all energy density components. The main contribution to
the gravitational potential comes from the DM perturbations which are
significantly altered by the interaction. The effect of the interaction
within the dark sector is propagated to other components by the
gravitational potential and, in particular, it changes
the peculiar velocity of baryons.

To compare with observations we compute the root mean square velocity
dispersion of the baryon velocity field, smoothed on a sphere of radius $r$,
defined as
\begin{equation}
\langle v_{\mathrm{b}}^2 \rangle = \int d^3k W_r^2(k) P_v(k),
\label{eq:peculiar}
\end{equation}
where $W_r(k)$ is a top hat window function
of radius $r$ and $P_v(k)$ is the power spectrum
of the baryon velocity field $v_{\mathrm{b}}$. The magnitude
$\langle v_{\mathrm{b}}^2 \rangle^{1/2}$ represents the mean velocity
of the matter within a sphere of radius $r$ with respect
to the mean matter distribution, usually known as bulk flow.
For comparison, we also compute the same magnitude
for the $\Lambda$CDM model.

Our results are presented in Figs.~\ref{x2} and
\ref{x1}. In the left panels we show the rms
velocity dispersion for spheres of different
radii and in the right panels the evolution of
gravitational potential for $k=1.0$Mpc$^{-1}$.
Lines follow the same convention in both left and
right panels. In Fig.~\ref{x2} the results are given for
$\xi_1=0$ and in Fig.~\ref{x1} for $\xi_2=0$.
These figures show that without interaction, the
gravitational potential falls slower with
decreasing EoS parameter $w$ at small
redshifts. This behavior leads to the
suppression of the rms peculiar velocity of
baryons for $w>-1$ and its enhancement for $w<-1$
compared with the $\Lambda$CDM model
($w=-1$). In interacting models the potential
evolves for a longer period than in
non-interacting models. If $\xi_1=0$, increasing the
coupling $\xi_2$ decreases the amplitude of the gravitational
potential and reduces the peculiar velocity compared
with the concordance model. For $\xi_2=0$,
the result is similar since larger $\xi_1$ gives smaller
peculiar velocities. Intuitively, stronger the coupling larger
the fraction of DE in the past, reducing the
amplitude of the gravitational field and consequently,
also the amplitude of the matter peculiar velocity.
For a positive coupling, the effect on the
peculiar velocities is larger when the coupling
is proportional to the DM than to the DE, as it
could be expected since the DM energy density is
larger than that of the DE during most of the
evolution history of the Universe. In Fig.~\ref{v_xt}
we show the peculiar velocity when the interaction
is proportional to the total energy density of the
dark sector. Again, the behavior is
very similar to the case with interaction
proportional to the DM energy only.
The greatest variation with respect to the concordance
model occurs when $\xi_2<0$ and
$w<-1$, when the peculiar velocity could be
larger by a factor of two.

Observationally it is difficult to measure peculiar
velocities on scales above $50h^{-1}$Mpc
using galaxies. In \cite{Kashlinsky2000} it was proposed
to use clusters as tracers of the velocity field since
the kSZ effect provides their peculiar velocity with
respect to the matter rest frame. The results of
\cite{Kashlinsky2008,Kashlinsky2010,Kashlinsky2011} suggest
that the local CMB dipole is not associated with our peculiar
motion but is intrinsic to the Last Scattering Surface.
If so, those results would not constrain any possible
interaction on the dark sector. A more direct probe would be the
measurement of the pair-wise velocity dispersion of
clusters \cite{Hand2012} since it is related to the
matter peculiar velocity. While the results from
Atacama Cosmology Telescope (ACT) \cite{Hand2012}
and {\it Planck} \cite{PlanckCollaboration2013a} seem
to be consistent with the $\Lambda$CDM model,
given their large uncertainties they are
also compatible with the models considered here. To
conclude, while at present the data does not have
the statistical power to constrain the interaction,
the peculiar velocity field could become
an important test of the interaction with future data sets
of higher resolution and lower noise.

\section{The Power Spectrum of Kinetic Sunyaev-Zel'dovich Temperature
Anisotropies}

The interaction changes the matter velocity field
at all redshifts, therefore it does not only affect the peculiar
velocity field on the local Universe, traced by
galaxies or clusters, but also the
velocity field since recombination. Therefore,
temperature anisotropies induced by the ionized
gas in unresolved sources up to the present will
generate a different pattern of kSZ temperature
anisotropies on the CMB and the power spectrum of
the kSZ effect will be sensitive to any DM-DE
interaction.

\begin{figure}[tbp]
\subfloat[]{
    \includegraphics[width=0.45\textwidth]{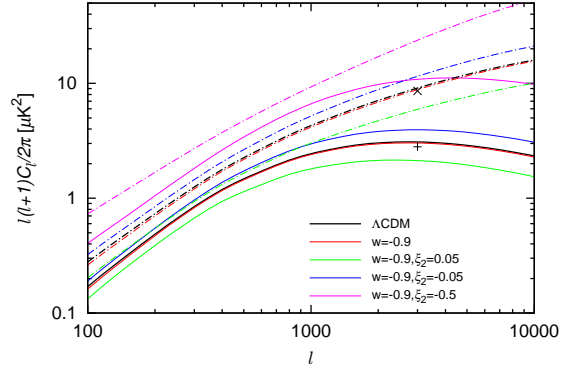}
    \label{wp1x2_ksz}
}\\
\subfloat[]{
    \includegraphics[width=0.45\textwidth]{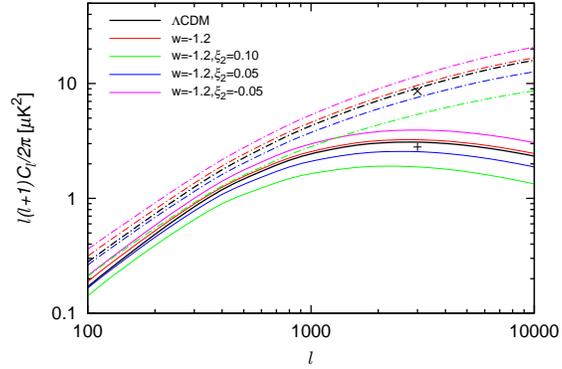}
    \label{wn1x2_ksz}
}\\
\subfloat[]{
    \includegraphics[width=0.45\textwidth]{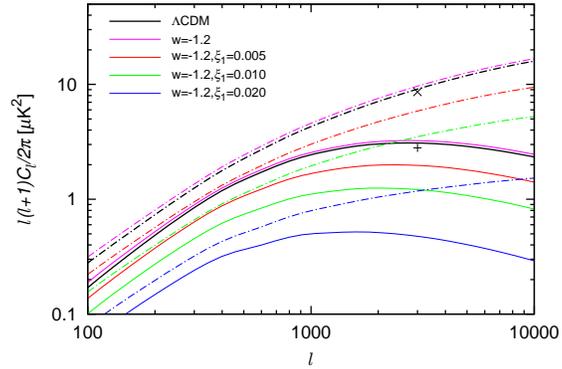}
    \label{wn1x1_ksz}
}
\caption{
Radiation power spectrum of the CMB temperature anisotropies
generated by the kinetic Sunyaev-Zel'dovich effect for different
interaction kernels: (a) $\xi_1=0,\; w>-1$; (b) $\xi_1=0,\; w<-1$ and
(c) $\xi_2=0,\; w<-1$. Solid lines correspond to the linear power spectra
and dot-dashed lines to non-linear power spectra.}
\label{ksz}
\end{figure}

To take into account the contribution of all the
ionized gas to the temperature anisotropies of
the CMB, we need to compute the effect of all
unresolved sources present in the Universe. At
redshift $z \lsim 10$, the reionization of the
intergalactic medium will produce free electrons
that would share the same motion as the average
baryon. CMB photons, re-scattered by this moving plasma,
give rise to secondary temperature anisotropies as \cite{Sunyaev1972}
\begin{equation}
\frac{\Delta T(\hat{\bm n})}{T_{\mathrm{CMB}}} =
-\int_{t_{\mathrm{re}}}^{t_0} n_{\mathrm{e}}\sigma_Te^{-\kappa}(\bm{v}\cdot\hat{\bm n})\mathrm{d}t,
\end{equation}
where $n_{\mathrm{e}}$ is the electron density, $\sigma_T$ is the Thomson
cross section and $\kappa$ is the Thomson optical depth, $\bm{v}$ is
the peculiar velocity of the electrons; the integral is along the
line of sight (l.o.s.) out to the reionization epoch and $\hat{\bm n}$ is
the unit vector along the l.o.s. Using the
comoving distance $x$ and neglecting any interaction of electrons
with other particles, we can write
$n_{\mathrm{e}}(x,z)=\chi_{\mathrm{e}}\bar{n}_{\mathrm{e}}(0)a^{-3}
[1+\delta_{\mathrm{e}}(x,z)]$,
where $\bar{n}_{\mathrm{e}}(0)$ is the mean electron number
density at present and $\chi_{\mathrm{e}}$ is the ionization
fraction. Within this approximation
\begin{equation}
\frac{\Delta T(\hat{\bm n})}{T_{\mathrm{CMB}}} =
\bar{n}_{\mathrm{e}}(0)\sigma_T\int a^{-2} \chi_{\mathrm{e}} e^{-\kappa}(\bm{p}\cdot\hat{\bm n})\mathrm{d}x,
\end{equation}
where we have defined the peculiar momentum
$\bm{p}\equiv (1+\delta_{\mathrm{e}})\bm{v}$;
$\bm{p}$ can be decomposed into a gradient component
$\bm{p}_{\mathrm{E}}$ and a curl component $\bm{p}_{\mathrm{B}}$.
The gradient term cancels out when integrating along the l.o.s. and has
no contribution to the kSZ effect. The net effect is
only due to the curl part of the peculiar momentum
\cite{Vishniac1987, Zhang2004}.

Since the kSZ effect induces temperature fluctuations at small
angular scales, the correlation function of the temperature
anisotropies between two positions in the sky separated by an
angle $\theta$ can be calculated using Limber approximation; then
\begin{equation}\begin{split}
w(\theta) &\approx (n_{\mathrm{e}}(0)\sigma_T)^2 \cos(\theta) \int^{x_{\mathrm{re}}}_0 \mathrm{d}x
(a^{-2}\chi_{\mathrm{e}})^2 \exp(-2\kappa(z)) \int^{\infty}_{-\infty}
\frac{1}{2}\xi_{\mathrm{B}}(\sqrt{x^2\theta^2+y^2}) dy \\
&\approx (n_{\mathrm{e}}(0)\sigma_T)^2 \int^{x_{\mathrm{re}}}_0 \mathrm{d}x (a^{-2}\chi_{\mathrm{e}})^2
\exp(-2\kappa(z)) \int^{\infty}_{-\infty} \frac{1}{2}\xi_{\mathrm{B}}(\sqrt{x^2\theta^2+y^2}) dy,
\end{split}
\label{eq:correlation}
\end{equation}
where $\xi_{\mathrm{B}}$ is the correlation function of ${\bm p}_{\mathrm{B}}$.
In eq.~(\ref{eq:correlation}) we have used the identity
$\langle\lvert {\bm p}_{\mathrm{B}}({\bm k})
\cdot \hat{\bm n} \rvert^2\rangle = P^2_{\mathrm{B}}(k)/2$, where $P_{\mathrm{B}}(k)$ is
the power spectrum of ${\bm p}_{\mathrm{B}}({\bm k})$. The contribution of
the kSZ effect to CMB temperature anisotropies now reads
\begin{equation}
C_l^{kSZ}=\frac{16\pi^2}{(2l+1)^3} (\bar{n}_{\mathrm{e}}(0)\sigma_T)^2
\int^{z_{\mathrm{re}}}_0 (1+z)^4 \chi^2_{\mathrm{e}} \frac{1}{2}\Delta^2_{\mathrm{B}}(k,z)\rvert_{k=l/x}
e^{-\kappa} x(z) \frac{dx(z)}{dz}dz,
\label{eq:cl}
\end{equation}
where $\Delta_{\mathrm{B}}^2(k,z) = \frac{k^3}{2\pi^2}P_{\mathrm{B}}(k,z)$.

Let us now analyze the contributions to the
solenoidal part of the electron velocity field.
Since $\nabla\times\bm{p} = (1+\delta_{\mathrm{e}})\nabla \times
\bm{v} + \nabla\delta_{\mathrm{e}} \times \bm{v}$, there are only two
possible contributions:
from the rotational mode of $\bm{v}$ or from the correlation between
the density and the velocity fields. In the linear regime, only the
irrotational component of the velocity fields couples to gravity,
so the first contribution is zero; the kSZ effect is due to the correlation
between the density gradient and the velocity.  In Fourier space,
\begin{equation}\begin{split}
P_{\mathrm{B}}(k) = \langle\bm{p}_{\mathrm{B}}^*(\bm{k})\bm{p}_{\mathrm{B}}(\tilde{\bm{k}})\rangle =
&\int\frac{\mathrm{d}^3\bm{k}'}{(2\pi)^3}
\int\frac{\mathrm{d}^3\tilde{\bm{k}}'}{(2\pi)^3}
\langle \delta_{\mathrm{e}}^*(\bm{k}-\bm{k}')\delta_{\mathrm{e}}(\tilde{\bm{k}}-\tilde{\bm{k}}')
\bm{v}^*(\bm{k}')\bm{v}(\tilde{\bm{k}}') \rangle \\
&\times \lvert \bm{k}' \rvert \lvert \tilde{\bm{k}}'
\rvert \beta(\bm{k},\bm{k}')\beta(\tilde{\bm{k}},\tilde{\bm{k}}'),
\end{split}
\label{pB}
\end{equation}
where
$\beta(\bm{k},\bm{k}') = [\bm{k}'-\bm{k}(\bm{k}\cdot\bm{k}')/\bm{k}^2]/\bm{k}'^2$.
Notice that in (\ref{pB}), $\bm{k'}$, $\tilde{\bm{k}}'$ and $\bm{k}-\bm{k'}$,
$\tilde{\bm{k}}-\tilde{\bm{k}}'$ can be interchanged.

In the linear perturbation theory, using the
subhorizon approximation, the relation between peculiar velocities and
density perturbations are given by
\begin{equation}
v=-\delta_{\mathrm{e}}'/k=-aHf_{\mathrm{e}}(a)\delta_{\mathrm{e}}/k,
\end{equation}
where $f_{\mathrm{e}}=\frac{d\ln \delta_{\mathrm{e}}}{d\ln a}$ is the
growth factor for electron density perturbations,
which at this order in perturbation theory is the
same as the baryonic matter growth
factor, but is different from that of
DM in the presence of DM-DE interaction. By defining $D_{\mathrm{e}}(z) \equiv
\delta_{\mathrm{e}}(z)/\delta_{\mathrm{e}}(0)$, we can rewrite
$aHf_{\mathrm{e}}=a\dot{D_{\mathrm{e}}}/{D_{\mathrm{e}}}$ and
\begin{equation}\begin{split}
P_{\mathrm{B}}(k,z) =
& \frac{a^2}{2} \int \frac{\mathrm{d}^3\bm{k}'}{(2\pi)^3}
\bigg(\frac{\dot{D_{\mathrm{e}}}}{D_{\mathrm{e}}}\bigg)^2 P(k',z) P(k-k',z)\\
&\times [W_g(k-k')\beta(\bm{k},\bm{k}') + W_g(k')\beta(\bm{k},\bm{k}-\bm{k}')]^2,
\end{split}
\label{P_B_e}
\end{equation}
where $P(k)$ is the baryon  power spectrum,
$W_g(k)$ is the transfer function which takes
into account the suppression of baryon density
fluctuations at small scales due to physical
processes \cite{Fang1993}.
For simplicity, we have set it to unity in
our numerical calculations.

Eqs.~(\ref{eq:cl},\ref{P_B_e}) give the
contributions of the ionized gas to the
temperature anisotropies of the CMB.
Let us remark that eq.~(\ref{eq:cl}) depends implicitly
on the DM-DE interaction. Since the fraction of ionized gas evolves as
\begin{equation}
\delta_{\mathrm{e}}''+\frac{a'}{a}\delta_{\mathrm{e}}'+3\phi''+3H\phi'+k^2\psi=0.
\end{equation}
In the small scale approximation the time variation of the potential can
be neglected with respect to their spatial gradients. Using Poisson equation
we can write
\begin{equation}
\delta_{\mathrm{e}}''+\frac{a'}{a}\delta_{\mathrm{e}}'-4\pi Ga^2(\rho_{\mathrm{b}}\delta_{\mathrm{b}}+\rho_{\mathrm{c}}\delta_{\mathrm{c}}+\rho_{\mathrm{d}}\delta_{\mathrm{d}})=0,
\end{equation}
Since DM and DE density perturbations depend on
the interaction (see
eqs.~\ref{eq:perturbations1}-\ref{eq:perturbations4}),
the time evolution of background and perturbed
magnitudes $\rho_{\mathrm{c}}$, $\rho_{\mathrm{d}}$, $\delta_{\mathrm{c}}$ and
$\delta_{\mathrm{d}}$ will be modified, changing the
evolution of the gravitational potential $\psi$.
The density perturbation of free electrons,
$\delta_{\mathrm{e}}$, will also be modified.
Due to the interaction, the evolution of both the
gravitational potential and density perturbations are scale-dependent,
which is different from that in the $\Lambda$CDM cosmology,
where all sub-horizon perturbations grow at the same rate.
Thus, a scale dependent behavior could be very useful to constrain
 observationally any possible interaction.

The non-linear evolution of matter density perturbations enhances
the kSZ effect. Following \cite{Hu2000, Ma2002, Shaw2011}, this
contribution changes the expression of (\ref{P_B_e}) to
\begin{equation}\begin{split}
P_{\mathrm{B}}(k,z) =
& \frac{a^2}{2} \int \frac{\mathrm{d}^3\bm{k}'}{(2\pi)^3}
\bigg(\frac{\dot{D_{\mathrm{e}}}}{D_{\mathrm{e}}}\bigg)^2 P(k',z) P(k-k',z)\\
&\times [W_g(k-k')T_{NL}(k-k')\beta(\bm{k},\bm{k}')
    + W_g(k')T_{NL}(k')\beta(\bm{k},\bm{k}-\bm{k}')]^2,
\end{split}
\label{P_B_e_NL}
\end{equation}
where we have defined the non-linear power spectrum
as $P^{NL}(k) \equiv P(k)T_{NL}^2(k)$. The nonlinear
correction affects the density and not the velocity field
\cite{Hu2000, Ma2002}, as could be expected since matter density
perturbations can be highly non-linear while the velocity
field is still in the linear regime \cite{Coles1991}.
To include the non-linear correction we need to specify
the non-linear power spectrum of baryon density perturbations
which is usually done by using adequate fits to
numerical simulations. Such simulations had not been carried
out for interacting DM and DE models so for this
type of models we can only guess what
the non-linear power spectrum could be,
making it difficult to give a reliable
estimate of the non-linear contribution. One further complication is
that on small scales the dynamics of baryons and pressureless DM
particles is very different. While non-linear corrections
can enhance the amplitude of the kSZ effect by a factor
2-4 in the range $\ell=3 \times 10^3 \sim 10^4$,
physical processes on the baryon component such as shock heating and
dissipation can reduce the change to
only 0.3-2, in the same range of multipoles \cite{Shaw2011}.

An estimate of the non-linear contribution can be given
if we assume that the nonlinear evolution of baryon density
perturbations is weakly dependent on the effect of the interaction.
Thus, we can construct the non-linear correction to the linear power
spectrum in the interacting model like in the concordance model:
$P^{NL}(k)=P(k)T_{NL}^2(k)$, where $P(k)$ is the baryon linear density
power spectrum, and the transfer function $T_{NL}(k)$ identical
to the concordance model. In the following numerical calculations,
we employ the halofit model developed in \cite{Smith2003} and revised
in \cite{Takahashi2012} to estimate the non-linear corrections.

In  Fig.~\ref{ksz} we represent the linear (solid lines)
and non-linear (dot-dashed lines) kSZ power
spectrum. The three plots correspond to our fiducial models:
(a) $\xi_1=0, \, w>-1$; (b) $\xi_1=0, \, w<-1$;
and (c) $\xi_2=0, \, w<-1$. The curves corresponding to the
non-linear kSZ effect are essentially upper limits since we do
not take into account the small scale smoothing of the baryon
density perturbations due to gas pressure.
Fig.~\ref{ksz} shows that in models with no
interaction the kSZ power spectrum is weakly
dependent on $w$. With DM-DE interaction, the
amplitude increases with decreasing $\xi_2$; it
is smaller than the non-interaction model when
$\xi_2>0$ and larger when $\xi_2<0$. In
Fig.~\ref{wn1x2_ksz}
the kSZ power spectrum amplitude also increases
with decreasing $\xi_2$. The same behavior is
observed in Fig.~\ref{wn1x1_ksz} with the
exception that the spectrum is always below
models without interaction since we have
restricted our study to the case $\xi_1>0$.
The results presented in Fig.~\ref{ksz}
follow those of Figs.~\ref{x2}, \ref{x1} that also
showed that positive couplings suppress the
amplitude of the peculiar velocities, resulting
in a smaller kSZ effect. Even though the kSZ spectrum is
enhanced by including the non-linear effect, the
scaling behavior with the coupling parameter is
the same as in the linear case.

In Fig.~\ref{ksz}, we include two
upper limits to the amplitude of the kSZ power spectrum at the
95\% confidence level. The cross, located at $8.6\mu K^2$
was derived from the ACT data at
$\ell=3000$ \cite{Sievers2013}. The plus
is the $2.8\mu K^2$ upper limit at
$\ell=3000$ derived from the South Pole Telescope (SPT) data
\cite{Reichardt2011}. These upper limits depend
on the reionization history and on the modeling
of Cosmic Infrared Background and thermal Sunyaev-Zel'dovich contributions. Nevertheless, these
results are already in tension with negative
coupling in DM-DE interaction and tend to favor
positive coupling (see Fig.~\ref{ksz}). A number of
complexities could change the constraints.
On the one hand, the kSZ effect from patchy
reionization is expected to be comparable with
that from the homogeneous part. Hence the $95\%$
upper limit on the homogeneous kSZ effect would be
significantly smaller than the quoted figures from
ACT and SPT. On the other, we did not include effect
of gas pressure in our calculation that could potentially
reduce the amplitude of the kSZ spectrum. Despite these
uncertainties, taken the data at face value, the
kSZ observations favor a positive coupling
between DM-DE interaction. This result is
encouraging, since it is consistent with our
previous constraints from the CMB \cite{He2009,
He2010a} and galaxy clusters \cite{Abdalla2007,
Abdalla2009}. Such positive coupling means that
the energy flows from DE to DM, as required to
alleviate the coincidence problem and to satisfy
the second law of thermodynamics \cite{Pavon2007}.
A positive coupling is also very reassuring in
the light of the coincidence problem. As Fig.~\ref{ratio}
shows, the background evolution contains a long period
in the expansion of the Universe where
DM and DE densities are comparable to each other.

\section{Conclusions}

In this paper we have studied the effect of the
DM-DE interaction in the gravitational field and
matter peculiar velocities.
We have analyzed how three different observables, the ISW
effect, the matter peculiar velocity and the CMB
temperature anisotropies induced by the kSZ
effect, are sensitive to the interaction.
The latter two effects have been considered in this
paper for the first time.
We have shown that the amplitude of
the kSZ effect, like the ISW effect, is sensitive
to the amplitude and sign of the interaction.
Although both effects are related to the
gravitational potential, the late-time ISW effect is
determined by the time-evolving gravitational
potential, while peculiar velocities and the kSZ
effect depend on the accelerations generated by
the potentials themselves. Both tests are complementary
with each other since they reflect physical processes
in the CMB that operate at different scales. For instance,
the ISW effect probes the effect of the interaction during the
period of accelerated expansion since the potentials are
roughly constant during the matter dominated regime while the
kSZ effect receives contribution since reionization at $z\lsim 10$.
Then, the time evolution of the gravitational potential that generate
the ISW effect contributes to CMB anisotropies mostly
at large angular scales while the kSZ effect, that depends
on the projection of the peculiar velocity field along
the line of sight, contributes preferentially at $\ell>10^3$.
Figs.~\ref{x2}, \ref{x1} show that even if
the evolution of the gravitational potential
changes little with time, the amplitude of the
gravitational potential can vary significantly due
to a DM-DE interaction. The
strong dependence of $\delta_{\mathrm{e}}$ and $v$,
which contributes to the rotational part of the
peculiar momentum, on the gravitational potential suggests that the
kSZ effect is more sensitive to the interaction than the ISW
effect. Furthermore, since the kSZ effect is
dominant at small angular scales, it is less
affected by cosmic variance. These secondary
anisotropies could become a powerful tool to
discriminate models with interaction.

By using the upper limits on the kSZ effect
from ACT and SPT, we have seen that these data
favors an interaction between DE and DM with a
positive coupling, which is consistent with the
previous CMB constraints \cite{He2009, He2010a}
and the galaxy cluster scale tests
\cite{Abdalla2007, Abdalla2009}. This positive
coupling means that there is energy flow from DE
to DM, which can alleviate the coincidence
problem. Further details of the kSZ spectrum
could be detected with the next generation of CMB
observations with high resolution, like
Gismo\footnote{http://www.iram.es/IRAMES/mainWiki/GoddardIramSuperconductingTwoMillimeterCamera}.
If this forthcoming experiment achieves the
expected sensitivity, it could provide the first
detection of the interaction within the dark
sector.

Alternatively, the peculiar velocity of resolved
clusters can be used to trace the velocity field
at different scales \cite{Kashlinsky2008,Kashlinsky2010,
Kashlinsky2011,Hand2012,Mroczkowski2012}. Models
with interaction could give peculiar velocities
with an amplitude about a factor of 2 larger or 5
smaller than the $\Lambda$CDM prediction. Combining
this observable with the ISW and kSZ power
spectrum can provide a measurement of the
interaction or upper limits on the amplitude of
the coupling stronger than those currently
available.

\begin{acknowledgments}
This work was partially supported by the National
Natural Science Foundation of China
and project FIS2012-30926 from the spanish
Ministerio de Ciencia e Innovaci\'on. We would
like to acknowledge helpful discussions with E.
Abdalla and D. Pavon.
\end{acknowledgments}

\bibliography{draft}

\begin{thebibliography}{80}
\expandafter\ifx\csname natexlab\endcsname\relax\def\natexlab#1{#1}\fi
\expandafter\ifx\csname bibnamefont\endcsname\relax
  \def\bibnamefont#1{#1}\fi
\expandafter\ifx\csname bibfnamefont\endcsname\relax
  \def\bibfnamefont#1{#1}\fi
\expandafter\ifx\csname citenamefont\endcsname\relax
  \def\citenamefont#1{#1}\fi
\expandafter\ifx\csname url\endcsname\relax
  \def\url#1{\texttt{#1}}\fi
\expandafter\ifx\csname urlprefix\endcsname\relax\def\urlprefix{URL }\fi
\providecommand{\bibinfo}[2]{#2}
\providecommand{\eprint}[2][]{\url{#2}}

\bibitem[{\citenamefont{Perlmutter et~al.}(1998)\citenamefont{Perlmutter,
  Aldering, Valle, Deustua, Ellis, Fabbro, Fruchter, Goldhaber, Goobar, Groom
  et~al.}}]{Perlmutter1998}
\bibinfo{author}{\bibfnamefont{S.}~\bibnamefont{Perlmutter}},
  \bibinfo{author}{\bibfnamefont{G.}~\bibnamefont{Aldering}},
  \bibinfo{author}{\bibfnamefont{M.~D.} \bibnamefont{Valle}},
  \bibinfo{author}{\bibfnamefont{S.}~\bibnamefont{Deustua}},
  \bibinfo{author}{\bibfnamefont{R.~S.} \bibnamefont{Ellis}},
  \bibinfo{author}{\bibfnamefont{S.}~\bibnamefont{Fabbro}},
  \bibinfo{author}{\bibfnamefont{A.}~\bibnamefont{Fruchter}},
  \bibinfo{author}{\bibfnamefont{G.}~\bibnamefont{Goldhaber}},
  \bibinfo{author}{\bibfnamefont{A.}~\bibnamefont{Goobar}},
  \bibinfo{author}{\bibfnamefont{D.~E.} \bibnamefont{Groom}},
  \bibnamefont{et~al.}, \bibinfo{journal}{Nature}
  \textbf{\bibinfo{volume}{391}}, \bibinfo{pages}{51} (\bibinfo{year}{1998}).

\bibitem[{\citenamefont{Riess et~al.}(1998)\citenamefont{Riess, Filippenko,
  Challis, Clocchiattia, Diercks, Garnavich, Gilliland, Hogan, Jha, Kirshner
  et~al.}}]{Riess1998}
\bibinfo{author}{\bibfnamefont{A.~G.} \bibnamefont{Riess}},
  \bibinfo{author}{\bibfnamefont{A.~V.} \bibnamefont{Filippenko}},
  \bibinfo{author}{\bibfnamefont{P.}~\bibnamefont{Challis}},
  \bibinfo{author}{\bibfnamefont{A.}~\bibnamefont{Clocchiattia}},
  \bibinfo{author}{\bibfnamefont{A.}~\bibnamefont{Diercks}},
  \bibinfo{author}{\bibfnamefont{P.~M.} \bibnamefont{Garnavich}},
  \bibinfo{author}{\bibfnamefont{R.~L.} \bibnamefont{Gilliland}},
  \bibinfo{author}{\bibfnamefont{C.~J.} \bibnamefont{Hogan}},
  \bibinfo{author}{\bibfnamefont{S.}~\bibnamefont{Jha}},
  \bibinfo{author}{\bibfnamefont{R.~P.} \bibnamefont{Kirshner}},
  \bibnamefont{et~al.}, \bibinfo{journal}{Astron.J.}
  \textbf{\bibinfo{volume}{116}}, \bibinfo{pages}{1009} (\bibinfo{year}{1998}).

\bibitem[{\citenamefont{Perlmutter et~al.}(1999)\citenamefont{Perlmutter,
  Aldering, Goldhaber, Knop, Nugent, Castro, Deustua, Fabbro, Goobar, Groom
  et~al.}}]{Perlmutter1999}
\bibinfo{author}{\bibfnamefont{S.}~\bibnamefont{Perlmutter}},
  \bibinfo{author}{\bibfnamefont{G.}~\bibnamefont{Aldering}},
  \bibinfo{author}{\bibfnamefont{G.}~\bibnamefont{Goldhaber}},
  \bibinfo{author}{\bibfnamefont{R.~A.} \bibnamefont{Knop}},
  \bibinfo{author}{\bibfnamefont{P.}~\bibnamefont{Nugent}},
  \bibinfo{author}{\bibfnamefont{P.~G.} \bibnamefont{Castro}},
  \bibinfo{author}{\bibfnamefont{S.}~\bibnamefont{Deustua}},
  \bibinfo{author}{\bibfnamefont{S.}~\bibnamefont{Fabbro}},
  \bibinfo{author}{\bibfnamefont{A.}~\bibnamefont{Goobar}},
  \bibinfo{author}{\bibfnamefont{D.~E.} \bibnamefont{Groom}},
  \bibnamefont{et~al.}, \bibinfo{journal}{Astrophys.J.}
  \textbf{\bibinfo{volume}{517}}, \bibinfo{pages}{565} (\bibinfo{year}{1999}).

\bibitem[{\citenamefont{Tonry et~al.}(2003)\citenamefont{Tonry, Schmidt,
  Barris, Candia, Challis, Clocchiatti, Coil, Filippenko, Garnavich, Hogan
  et~al.}}]{Tonry2003}
\bibinfo{author}{\bibfnamefont{J.~L.} \bibnamefont{Tonry}},
  \bibinfo{author}{\bibfnamefont{B.~P.} \bibnamefont{Schmidt}},
  \bibinfo{author}{\bibfnamefont{B.}~\bibnamefont{Barris}},
  \bibinfo{author}{\bibfnamefont{P.}~\bibnamefont{Candia}},
  \bibinfo{author}{\bibfnamefont{P.}~\bibnamefont{Challis}},
  \bibinfo{author}{\bibfnamefont{A.}~\bibnamefont{Clocchiatti}},
  \bibinfo{author}{\bibfnamefont{A.~L.} \bibnamefont{Coil}},
  \bibinfo{author}{\bibfnamefont{A.~V.} \bibnamefont{Filippenko}},
  \bibinfo{author}{\bibfnamefont{P.}~\bibnamefont{Garnavich}},
  \bibinfo{author}{\bibfnamefont{C.}~\bibnamefont{Hogan}},
  \bibnamefont{et~al.}, \bibinfo{journal}{Astrophys.J.}
  \textbf{\bibinfo{volume}{594}}, \bibinfo{pages}{1} (\bibinfo{year}{2003}).

\bibitem[{\citenamefont{Riess et~al.}(2004)\citenamefont{Riess, Strolger,
  Tonry, Casertano, Ferguson, Mobasher, Challis, Filippenko, Jha, Li
  et~al.}}]{Riess2004}
\bibinfo{author}{\bibfnamefont{A.~G.} \bibnamefont{Riess}},
  \bibinfo{author}{\bibfnamefont{L.-G.} \bibnamefont{Strolger}},
  \bibinfo{author}{\bibfnamefont{J.}~\bibnamefont{Tonry}},
  \bibinfo{author}{\bibfnamefont{S.}~\bibnamefont{Casertano}},
  \bibinfo{author}{\bibfnamefont{H.~C.} \bibnamefont{Ferguson}},
  \bibinfo{author}{\bibfnamefont{B.}~\bibnamefont{Mobasher}},
  \bibinfo{author}{\bibfnamefont{P.}~\bibnamefont{Challis}},
  \bibinfo{author}{\bibfnamefont{A.~V.} \bibnamefont{Filippenko}},
  \bibinfo{author}{\bibfnamefont{S.}~\bibnamefont{Jha}},
  \bibinfo{author}{\bibfnamefont{W.}~\bibnamefont{Li}}, \bibnamefont{et~al.},
  \bibinfo{journal}{Astrophys.J.} \textbf{\bibinfo{volume}{607}},
  \bibinfo{pages}{665} (\bibinfo{year}{2004}).

\bibitem[{\citenamefont{Astier et~al.}(2006)\citenamefont{Astier, Guy,
  Regnault, Pain, Aubourg, Balam, Basa, Carlberg, Fabbro, Fouchez
  et~al.}}]{Astier2006}
\bibinfo{author}{\bibfnamefont{P.}~\bibnamefont{Astier}},
  \bibinfo{author}{\bibfnamefont{J.}~\bibnamefont{Guy}},
  \bibinfo{author}{\bibfnamefont{N.}~\bibnamefont{Regnault}},
  \bibinfo{author}{\bibfnamefont{R.}~\bibnamefont{Pain}},
  \bibinfo{author}{\bibfnamefont{E.}~\bibnamefont{Aubourg}},
  \bibinfo{author}{\bibfnamefont{D.}~\bibnamefont{Balam}},
  \bibinfo{author}{\bibfnamefont{S.}~\bibnamefont{Basa}},
  \bibinfo{author}{\bibfnamefont{R.~G.} \bibnamefont{Carlberg}},
  \bibinfo{author}{\bibfnamefont{S.}~\bibnamefont{Fabbro}},
  \bibinfo{author}{\bibfnamefont{D.}~\bibnamefont{Fouchez}},
  \bibnamefont{et~al.}, \bibinfo{journal}{Astron.Astrophys.}
  \textbf{\bibinfo{volume}{447}}, \bibinfo{pages}{31} (\bibinfo{year}{2006}).

\bibitem[{\citenamefont{Riess et~al.}(2007)\citenamefont{Riess, Strolger,
  Casertano, Ferguson, Mobasher, Gold, Challis, Filippenko, Jha, Li
  et~al.}}]{Riess2007}
\bibinfo{author}{\bibfnamefont{A.~G.} \bibnamefont{Riess}},
  \bibinfo{author}{\bibfnamefont{L.-G.} \bibnamefont{Strolger}},
  \bibinfo{author}{\bibfnamefont{S.}~\bibnamefont{Casertano}},
  \bibinfo{author}{\bibfnamefont{H.~C.} \bibnamefont{Ferguson}},
  \bibinfo{author}{\bibfnamefont{B.}~\bibnamefont{Mobasher}},
  \bibinfo{author}{\bibfnamefont{B.}~\bibnamefont{Gold}},
  \bibinfo{author}{\bibfnamefont{P.~J.} \bibnamefont{Challis}},
  \bibinfo{author}{\bibfnamefont{A.~V.} \bibnamefont{Filippenko}},
  \bibinfo{author}{\bibfnamefont{S.}~\bibnamefont{Jha}},
  \bibinfo{author}{\bibfnamefont{W.}~\bibnamefont{Li}}, \bibnamefont{et~al.},
  \bibinfo{journal}{Astrophys.J.} \textbf{\bibinfo{volume}{659}},
  \bibinfo{pages}{98} (\bibinfo{year}{2007}).

\bibitem[{\citenamefont{Hinshaw et~al.}(2012)\citenamefont{Hinshaw, Larson,
  Komatsu, Spergel, Bennett, Dunkley, Nolta, Halpern, Hill, Odegard
  et~al.}}]{Hinshaw2012}
\bibinfo{author}{\bibfnamefont{G.}~\bibnamefont{Hinshaw}},
  \bibinfo{author}{\bibfnamefont{D.}~\bibnamefont{Larson}},
  \bibinfo{author}{\bibfnamefont{E.}~\bibnamefont{Komatsu}},
  \bibinfo{author}{\bibfnamefont{D.~N.} \bibnamefont{Spergel}},
  \bibinfo{author}{\bibfnamefont{C.~L.} \bibnamefont{Bennett}},
  \bibinfo{author}{\bibfnamefont{J.}~\bibnamefont{Dunkley}},
  \bibinfo{author}{\bibfnamefont{M.~R.} \bibnamefont{Nolta}},
  \bibinfo{author}{\bibfnamefont{M.}~\bibnamefont{Halpern}},
  \bibinfo{author}{\bibfnamefont{R.~S.} \bibnamefont{Hill}},
  \bibinfo{author}{\bibfnamefont{N.}~\bibnamefont{Odegard}},
  \bibnamefont{et~al.}, \bibinfo{journal}{ApJS in press}
  (\bibinfo{year}{2012}).

\bibitem[{\citenamefont{PlanckCollaboration}(2013)}]{PlanckCollaboration2013b}
\bibinfo{author}{\bibnamefont{PlanckCollaboration}},
  \bibinfo{journal}{Submitted to A\&A}  (\bibinfo{year}{2013}).

\bibitem[{\citenamefont{Micheletti et~al.}(2009)\citenamefont{Micheletti,
  Abdalla, and Wang}}]{Micheletti2009}
\bibinfo{author}{\bibfnamefont{S.}~\bibnamefont{Micheletti}},
  \bibinfo{author}{\bibfnamefont{E.}~\bibnamefont{Abdalla}}, \bibnamefont{and}
  \bibinfo{author}{\bibfnamefont{B.}~\bibnamefont{Wang}},
  \bibinfo{journal}{Phys.Rev.D} \textbf{\bibinfo{volume}{79}},
  \bibinfo{pages}{123506} (\bibinfo{year}{2009}).

\bibitem[{\citenamefont{Micheletti}(2009)}]{Micheletti2009a}
\bibinfo{author}{\bibfnamefont{S.~M.~R.} \bibnamefont{Micheletti}},
  \bibinfo{journal}{JCAP} \textbf{\bibinfo{volume}{1005}}, \bibinfo{pages}{009}
  (\bibinfo{year}{2009}).

\bibitem[{\citenamefont{Amendola}(2000)}]{Amendola2000}
\bibinfo{author}{\bibfnamefont{L.}~\bibnamefont{Amendola}},
  \bibinfo{journal}{Phys.Rev.D} \textbf{\bibinfo{volume}{62}},
  \bibinfo{pages}{043511} (\bibinfo{year}{2000}).

\bibitem[{\citenamefont{Amendola and Quercellini}(2003)}]{Amendola2003}
\bibinfo{author}{\bibfnamefont{L.}~\bibnamefont{Amendola}} \bibnamefont{and}
  \bibinfo{author}{\bibfnamefont{C.}~\bibnamefont{Quercellini}},
  \bibinfo{journal}{Phys.Rev.D} \textbf{\bibinfo{volume}{68}},
  \bibinfo{pages}{023514} (\bibinfo{year}{2003}).

\bibitem[{\citenamefont{Amendola et~al.}(2006)\citenamefont{Amendola,
  Tsujikawa, and Sami}}]{Amendola2006}
\bibinfo{author}{\bibfnamefont{L.}~\bibnamefont{Amendola}},
  \bibinfo{author}{\bibfnamefont{S.}~\bibnamefont{Tsujikawa}},
  \bibnamefont{and} \bibinfo{author}{\bibfnamefont{M.}~\bibnamefont{Sami}},
  \bibinfo{journal}{Phys.Lett. B} \textbf{\bibinfo{volume}{632}},
  \bibinfo{pages}{155} (\bibinfo{year}{2006}).

\bibitem[{\citenamefont{Pavon and Zimdahl}(2005)}]{Pavon2005}
\bibinfo{author}{\bibfnamefont{D.}~\bibnamefont{Pavon}} \bibnamefont{and}
  \bibinfo{author}{\bibfnamefont{W.}~\bibnamefont{Zimdahl}},
  \bibinfo{journal}{Phys.Lett.B} \textbf{\bibinfo{volume}{628}},
  \bibinfo{pages}{206} (\bibinfo{year}{2005}).

\bibitem[{\citenamefont{del Campo et~al.}(2008)\citenamefont{del Campo,
  Herrera, and Pavon}}]{Campo2008}
\bibinfo{author}{\bibfnamefont{S.}~\bibnamefont{del Campo}},
  \bibinfo{author}{\bibfnamefont{R.}~\bibnamefont{Herrera}}, \bibnamefont{and}
  \bibinfo{author}{\bibfnamefont{D.}~\bibnamefont{Pavon}},
  \bibinfo{journal}{Phys.Rev.D} \textbf{\bibinfo{volume}{78}},
  \bibinfo{pages}{021302} (\bibinfo{year}{2008}).

\bibitem[{\citenamefont{Boehmer et~al.}(2008)\citenamefont{Boehmer,
  Caldera-Cabral, Lazkoz, and Maartens}}]{Boehmer2008}
\bibinfo{author}{\bibfnamefont{C.~G.} \bibnamefont{Boehmer}},
  \bibinfo{author}{\bibfnamefont{G.}~\bibnamefont{Caldera-Cabral}},
  \bibinfo{author}{\bibfnamefont{R.}~\bibnamefont{Lazkoz}}, \bibnamefont{and}
  \bibinfo{author}{\bibfnamefont{R.}~\bibnamefont{Maartens}},
  \bibinfo{journal}{Phys.Rev.D} \textbf{\bibinfo{volume}{78}},
  \bibinfo{pages}{023505} (\bibinfo{year}{2008}).

\bibitem[{\citenamefont{Chen et~al.}(2008)\citenamefont{Chen, Wang, and
  Jing}}]{Chen2008}
\bibinfo{author}{\bibfnamefont{S.}~\bibnamefont{Chen}},
  \bibinfo{author}{\bibfnamefont{B.}~\bibnamefont{Wang}}, \bibnamefont{and}
  \bibinfo{author}{\bibfnamefont{J.}~\bibnamefont{Jing}},
  \bibinfo{journal}{Phys.Rev.D} \textbf{\bibinfo{volume}{78}},
  \bibinfo{pages}{123503} (\bibinfo{year}{2008}).

\bibitem[{\citenamefont{Wang et~al.}(2005)\citenamefont{Wang, Gong, and
  Abdalla}}]{Wang2005}
\bibinfo{author}{\bibfnamefont{B.}~\bibnamefont{Wang}},
  \bibinfo{author}{\bibfnamefont{Y.}~\bibnamefont{Gong}}, \bibnamefont{and}
  \bibinfo{author}{\bibfnamefont{E.}~\bibnamefont{Abdalla}},
  \bibinfo{journal}{Physics Letters B} \textbf{\bibinfo{volume}{624}},
  \bibinfo{pages}{141 } (\bibinfo{year}{2005}).

\bibitem[{\citenamefont{Das et~al.}(2006)\citenamefont{Das, Corasaniti, and
  Khoury}}]{Das2006}
\bibinfo{author}{\bibfnamefont{S.}~\bibnamefont{Das}},
  \bibinfo{author}{\bibfnamefont{P.~S.} \bibnamefont{Corasaniti}},
  \bibnamefont{and} \bibinfo{author}{\bibfnamefont{J.}~\bibnamefont{Khoury}},
  \bibinfo{journal}{Phys. Rev. D} \textbf{\bibinfo{volume}{73}},
  \bibinfo{pages}{083509} (\bibinfo{year}{2006}).

\bibitem[{\citenamefont{Olivares et~al.}(2005)\citenamefont{Olivares,
  Atrio-Barandela, and Pav\'on}}]{Olivares2005}
\bibinfo{author}{\bibfnamefont{G.~a.~n.} \bibnamefont{Olivares}},
  \bibinfo{author}{\bibfnamefont{F.}~\bibnamefont{Atrio-Barandela}},
  \bibnamefont{and} \bibinfo{author}{\bibfnamefont{D.}~\bibnamefont{Pav\'on}},
  \bibinfo{journal}{Phys. Rev. D} \textbf{\bibinfo{volume}{71}},
  \bibinfo{pages}{063523} (\bibinfo{year}{2005}).

\bibitem[{\citenamefont{Olivares
  et~al.}(2008{\natexlab{a}})\citenamefont{Olivares, Atrio-Barandela, and
  Pav\'on}}]{Olivares2008a}
\bibinfo{author}{\bibfnamefont{G.}~\bibnamefont{Olivares}},
  \bibinfo{author}{\bibfnamefont{F.}~\bibnamefont{Atrio-Barandela}},
  \bibnamefont{and} \bibinfo{author}{\bibfnamefont{D.}~\bibnamefont{Pav\'on}},
  \bibinfo{journal}{Phys. Rev. D} \textbf{\bibinfo{volume}{77}},
  \bibinfo{pages}{063513} (\bibinfo{year}{2008}{\natexlab{a}}).

\bibitem[{\citenamefont{Valiviita et~al.}(2008)\citenamefont{Valiviita,
  Majerotto, and Maartens}}]{Valiviita2008}
\bibinfo{author}{\bibfnamefont{J.}~\bibnamefont{Valiviita}},
  \bibinfo{author}{\bibfnamefont{E.}~\bibnamefont{Majerotto}},
  \bibnamefont{and} \bibinfo{author}{\bibfnamefont{R.}~\bibnamefont{Maartens}},
  \bibinfo{journal}{JCAP} \textbf{\bibinfo{volume}{07}}, \bibinfo{pages}{020}
  (\bibinfo{year}{2008}).

\bibitem[{\citenamefont{He et~al.}(2008)\citenamefont{He, Wang, and
  Abdalla}}]{He2008a}
\bibinfo{author}{\bibfnamefont{J.-H.} \bibnamefont{He}},
  \bibinfo{author}{\bibfnamefont{B.}~\bibnamefont{Wang}}, \bibnamefont{and}
  \bibinfo{author}{\bibfnamefont{E.}~\bibnamefont{Abdalla}},
  \bibinfo{journal}{Phys.Lett.B} \textbf{\bibinfo{volume}{671}},
  \bibinfo{pages}{139} (\bibinfo{year}{2008}).

\bibitem[{\citenamefont{Corasaniti}(2008)}]{Corasaniti2008}
\bibinfo{author}{\bibfnamefont{P.~S.} \bibnamefont{Corasaniti}},
  \bibinfo{journal}{Phys.Rev.D} \textbf{\bibinfo{volume}{78}},
  \bibinfo{pages}{083538} (\bibinfo{year}{2008}).

\bibitem[{\citenamefont{Jackson et~al.}(2009)\citenamefont{Jackson, Taylor, and
  Berera}}]{Jackson2009}
\bibinfo{author}{\bibfnamefont{B.~M.} \bibnamefont{Jackson}},
  \bibinfo{author}{\bibfnamefont{A.}~\bibnamefont{Taylor}}, \bibnamefont{and}
  \bibinfo{author}{\bibfnamefont{A.}~\bibnamefont{Berera}},
  \bibinfo{journal}{Phys. Rev. D} \textbf{\bibinfo{volume}{79}},
  \bibinfo{pages}{043526} (\bibinfo{year}{2009}).

\bibitem[{\citenamefont{Pavon and Wang}(2007)}]{Pavon2007}
\bibinfo{author}{\bibfnamefont{D.}~\bibnamefont{Pavon}} \bibnamefont{and}
  \bibinfo{author}{\bibfnamefont{B.}~\bibnamefont{Wang}},
  \bibinfo{journal}{Gen.Rel.Grav.} \textbf{\bibinfo{volume}{41}},
  \bibinfo{pages}{1} (\bibinfo{year}{2007}).

\bibitem[{\citenamefont{Wang et~al.}(2007{\natexlab{a}})\citenamefont{Wang,
  Lin, Pavon, and Abdalla}}]{Wang2007}
\bibinfo{author}{\bibfnamefont{B.}~\bibnamefont{Wang}},
  \bibinfo{author}{\bibfnamefont{C.-Y.} \bibnamefont{Lin}},
  \bibinfo{author}{\bibfnamefont{D.}~\bibnamefont{Pavon}}, \bibnamefont{and}
  \bibinfo{author}{\bibfnamefont{E.}~\bibnamefont{Abdalla}},
  \bibinfo{journal}{Phys.Lett.B} \textbf{\bibinfo{volume}{662}},
  \bibinfo{pages}{1} (\bibinfo{year}{2007}{\natexlab{a}}).

\bibitem[{\citenamefont{Wang et~al.}(2007{\natexlab{b}})\citenamefont{Wang,
  Zang, Lin, Abdalla, and Micheletti}}]{Wang2007a}
\bibinfo{author}{\bibfnamefont{B.}~\bibnamefont{Wang}},
  \bibinfo{author}{\bibfnamefont{J.}~\bibnamefont{Zang}},
  \bibinfo{author}{\bibfnamefont{C.-Y.} \bibnamefont{Lin}},
  \bibinfo{author}{\bibfnamefont{E.}~\bibnamefont{Abdalla}}, \bibnamefont{and}
  \bibinfo{author}{\bibfnamefont{S.}~\bibnamefont{Micheletti}},
  \bibinfo{journal}{Nucl.Phys.B} \textbf{\bibinfo{volume}{778}},
  \bibinfo{pages}{69} (\bibinfo{year}{2007}{\natexlab{b}}).

\bibitem[{\citenamefont{Simpson et~al.}(2011)\citenamefont{Simpson, Jackson,
  and Peacock}}]{Simpson2010}
\bibinfo{author}{\bibfnamefont{F.}~\bibnamefont{Simpson}},
  \bibinfo{author}{\bibfnamefont{B.~M.} \bibnamefont{Jackson}},
  \bibnamefont{and} \bibinfo{author}{\bibfnamefont{J.~A.}
  \bibnamefont{Peacock}}, \bibinfo{journal}{MNRAS}
  \textbf{\bibinfo{volume}{411}}, \bibinfo{pages}{1053} (\bibinfo{year}{2011}).

\bibitem[{\citenamefont{Zimdahl}(2005)}]{Zimdahl2005}
\bibinfo{author}{\bibfnamefont{W.}~\bibnamefont{Zimdahl}},
  \bibinfo{journal}{Int.J.Mod.Phys. D} \textbf{\bibinfo{volume}{14}},
  \bibinfo{pages}{2319} (\bibinfo{year}{2005}).

\bibitem[{\citenamefont{Guo et~al.}(2007)\citenamefont{Guo, Ohta, and
  Tsujikawa}}]{Guo2007}
\bibinfo{author}{\bibfnamefont{Z.-K.} \bibnamefont{Guo}},
  \bibinfo{author}{\bibfnamefont{N.}~\bibnamefont{Ohta}}, \bibnamefont{and}
  \bibinfo{author}{\bibfnamefont{S.}~\bibnamefont{Tsujikawa}},
  \bibinfo{journal}{Phys.Rev.D} \textbf{\bibinfo{volume}{76}},
  \bibinfo{pages}{023508} (\bibinfo{year}{2007}).

\bibitem[{\citenamefont{Feng et~al.}(2008)\citenamefont{Feng, Wang, Abdalla,
  and Su}}]{Feng2008}
\bibinfo{author}{\bibfnamefont{C.}~\bibnamefont{Feng}},
  \bibinfo{author}{\bibfnamefont{B.}~\bibnamefont{Wang}},
  \bibinfo{author}{\bibfnamefont{E.}~\bibnamefont{Abdalla}}, \bibnamefont{and}
  \bibinfo{author}{\bibfnamefont{R.-K.} \bibnamefont{Su}},
  \bibinfo{journal}{Phys.Lett.B} \textbf{\bibinfo{volume}{665}},
  \bibinfo{pages}{111} (\bibinfo{year}{2008}).

\bibitem[{\citenamefont{Valiviita et~al.}(2009)\citenamefont{Valiviita,
  Maartens, and Majerotto}}]{Valiviita2009}
\bibinfo{author}{\bibfnamefont{J.}~\bibnamefont{Valiviita}},
  \bibinfo{author}{\bibfnamefont{R.}~\bibnamefont{Maartens}}, \bibnamefont{and}
  \bibinfo{author}{\bibfnamefont{E.}~\bibnamefont{Majerotto}},
  \bibinfo{journal}{Mon. Not. R. Astron. Soc.} \textbf{\bibinfo{volume}{402,}},
  \bibinfo{pages}{2355} (\bibinfo{year}{2009}).

\bibitem[{\citenamefont{Xia}(2009)}]{Xia2009}
\bibinfo{author}{\bibfnamefont{J.-Q.} \bibnamefont{Xia}},
  \bibinfo{journal}{Phys.Rev.D} \textbf{\bibinfo{volume}{80}},
  \bibinfo{pages}{103514} (\bibinfo{year}{2009}).

\bibitem[{\citenamefont{He et~al.}(2009{\natexlab{a}})\citenamefont{He, Wang,
  and Zhang}}]{He2009}
\bibinfo{author}{\bibfnamefont{J.-H.} \bibnamefont{He}},
  \bibinfo{author}{\bibfnamefont{B.}~\bibnamefont{Wang}}, \bibnamefont{and}
  \bibinfo{author}{\bibfnamefont{P.}~\bibnamefont{Zhang}},
  \bibinfo{journal}{Phys.Rev.D} \textbf{\bibinfo{volume}{80}},
  \bibinfo{pages}{063530} (\bibinfo{year}{2009}{\natexlab{a}}).

\bibitem[{\citenamefont{Martinelli et~al.}(2010)\citenamefont{Martinelli,
  Honorez, Melchiorri, and Mena}}]{Martinelli2010}
\bibinfo{author}{\bibfnamefont{M.}~\bibnamefont{Martinelli}},
  \bibinfo{author}{\bibfnamefont{L.~L.} \bibnamefont{Honorez}},
  \bibinfo{author}{\bibfnamefont{A.}~\bibnamefont{Melchiorri}},
  \bibnamefont{and} \bibinfo{author}{\bibfnamefont{O.}~\bibnamefont{Mena}},
  \bibinfo{journal}{Phys.Rev.D} \textbf{\bibinfo{volume}{81}},
  \bibinfo{pages}{103534} (\bibinfo{year}{2010}).

\bibitem[{\citenamefont{Honorez et~al.}(2010)\citenamefont{Honorez, Reid, Mena,
  Verde, and Jimenez}}]{Honorez2010a}
\bibinfo{author}{\bibfnamefont{L.~L.} \bibnamefont{Honorez}},
  \bibinfo{author}{\bibfnamefont{B.~A.} \bibnamefont{Reid}},
  \bibinfo{author}{\bibfnamefont{O.}~\bibnamefont{Mena}},
  \bibinfo{author}{\bibfnamefont{L.}~\bibnamefont{Verde}}, \bibnamefont{and}
  \bibinfo{author}{\bibfnamefont{R.}~\bibnamefont{Jimenez}},
  \bibinfo{journal}{JCAP} \textbf{\bibinfo{volume}{09}}, \bibinfo{pages}{029}
  (\bibinfo{year}{2010}).

\bibitem[{\citenamefont{He and Wang}(2008)}]{He2008}
\bibinfo{author}{\bibfnamefont{J.-H.} \bibnamefont{He}} \bibnamefont{and}
  \bibinfo{author}{\bibfnamefont{B.}~\bibnamefont{Wang}},
  \bibinfo{journal}{JCAP} \textbf{\bibinfo{volume}{06}}, \bibinfo{pages}{010}
  (\bibinfo{year}{2008}).

\bibitem[{\citenamefont{He et~al.}(2009{\natexlab{b}})\citenamefont{He, Wang,
  and Jing}}]{He2009a}
\bibinfo{author}{\bibfnamefont{J.-H.} \bibnamefont{He}},
  \bibinfo{author}{\bibfnamefont{B.}~\bibnamefont{Wang}}, \bibnamefont{and}
  \bibinfo{author}{\bibfnamefont{Y.~P.} \bibnamefont{Jing}},
  \bibinfo{journal}{JCAP} \textbf{\bibinfo{volume}{07}}, \bibinfo{pages}{030}
  (\bibinfo{year}{2009}{\natexlab{b}}).

\bibitem[{\citenamefont{Caldera-Cabral
  et~al.}(2009)\citenamefont{Caldera-Cabral, Maartens, and
  Schaefer}}]{Caldera-Cabral2009}
\bibinfo{author}{\bibfnamefont{G.}~\bibnamefont{Caldera-Cabral}},
  \bibinfo{author}{\bibfnamefont{R.}~\bibnamefont{Maartens}}, \bibnamefont{and}
  \bibinfo{author}{\bibfnamefont{B.~M.} \bibnamefont{Schaefer}},
  \bibinfo{journal}{JCAP} \textbf{\bibinfo{volume}{07}}, \bibinfo{pages}{027}
  (\bibinfo{year}{2009}).

\bibitem[{\citenamefont{He et~al.}(2010)\citenamefont{He, Wang, Abdalla, and
  Pavon}}]{He2010}
\bibinfo{author}{\bibfnamefont{J.-H.} \bibnamefont{He}},
  \bibinfo{author}{\bibfnamefont{B.}~\bibnamefont{Wang}},
  \bibinfo{author}{\bibfnamefont{E.}~\bibnamefont{Abdalla}}, \bibnamefont{and}
  \bibinfo{author}{\bibfnamefont{D.}~\bibnamefont{Pavon}},
  \bibinfo{journal}{JCAP} \textbf{\bibinfo{volume}{12}}, \bibinfo{pages}{022}
  (\bibinfo{year}{2010}).

\bibitem[{\citenamefont{Bertolami et~al.}(2007)\citenamefont{Bertolami, Pedro,
  and Delliou}}]{Bertolami2007}
\bibinfo{author}{\bibfnamefont{O.}~\bibnamefont{Bertolami}},
  \bibinfo{author}{\bibfnamefont{F.~G.} \bibnamefont{Pedro}}, \bibnamefont{and}
  \bibinfo{author}{\bibfnamefont{M.~L.} \bibnamefont{Delliou}},
  \bibinfo{journal}{Phys.Lett.B} \textbf{\bibinfo{volume}{654}},
  \bibinfo{pages}{165} (\bibinfo{year}{2007}).

\bibitem[{\citenamefont{Bertolami et~al.}(2009)\citenamefont{Bertolami,
  Gil~Pedro, and Le~Delliou}}]{Bertolami2009}
\bibinfo{author}{\bibfnamefont{O.}~\bibnamefont{Bertolami}},
  \bibinfo{author}{\bibfnamefont{F.}~\bibnamefont{Gil~Pedro}},
  \bibnamefont{and}
  \bibinfo{author}{\bibfnamefont{M.}~\bibnamefont{Le~Delliou}},
  \bibinfo{journal}{General Relativity and Gravitation}
  \textbf{\bibinfo{volume}{41}}, \bibinfo{pages}{2839} (\bibinfo{year}{2009}).

\bibitem[{\citenamefont{Abdalla
  et~al.}(2009{\natexlab{a}})\citenamefont{Abdalla, Abramo, Sodre, and
  Wang}}]{Abdalla2007}
\bibinfo{author}{\bibfnamefont{E.}~\bibnamefont{Abdalla}},
  \bibinfo{author}{\bibfnamefont{L.~R.} \bibnamefont{Abramo}},
  \bibinfo{author}{\bibfnamefont{L.}~\bibnamefont{Sodre}}, \bibnamefont{and}
  \bibinfo{author}{\bibfnamefont{B.}~\bibnamefont{Wang}},
  \bibinfo{journal}{Physics Letters B} \textbf{\bibinfo{volume}{673}},
  \bibinfo{pages}{107} (\bibinfo{year}{2009}{\natexlab{a}}).

\bibitem[{\citenamefont{Abdalla
  et~al.}(2009{\natexlab{b}})\citenamefont{Abdalla, Abramo, and
  de~Souza}}]{Abdalla2009}
\bibinfo{author}{\bibfnamefont{E.}~\bibnamefont{Abdalla}},
  \bibinfo{author}{\bibfnamefont{L.~R.} \bibnamefont{Abramo}},
  \bibnamefont{and} \bibinfo{author}{\bibfnamefont{J.~C.~C.}
  \bibnamefont{de~Souza}}, \bibinfo{journal}{Phys. Rev. D}
  \textbf{\bibinfo{volume}{82}}, \bibinfo{pages}{023508}
  (\bibinfo{year}{2009}{\natexlab{b}}).

\bibitem[{\citenamefont{Olivares et~al.}(2006)\citenamefont{Olivares,
  Atrio-Barandela, and Pavon}}]{Olivares2006}
\bibinfo{author}{\bibfnamefont{G.}~\bibnamefont{Olivares}},
  \bibinfo{author}{\bibfnamefont{F.}~\bibnamefont{Atrio-Barandela}},
  \bibnamefont{and} \bibinfo{author}{\bibfnamefont{D.}~\bibnamefont{Pavon}},
  \bibinfo{journal}{Phys.Rev.D} \textbf{\bibinfo{volume}{74}},
  \bibinfo{pages}{043521} (\bibinfo{year}{2006}).

\bibitem[{\citenamefont{He et~al.}(2011)\citenamefont{He, Wang, and
  Abdalla}}]{He2010a}
\bibinfo{author}{\bibfnamefont{J.-H.} \bibnamefont{He}},
  \bibinfo{author}{\bibfnamefont{B.}~\bibnamefont{Wang}}, \bibnamefont{and}
  \bibinfo{author}{\bibfnamefont{E.}~\bibnamefont{Abdalla}},
  \bibinfo{journal}{Phys.Rev.D} \textbf{\bibinfo{volume}{83}},
  \bibinfo{pages}{063515} (\bibinfo{year}{2011}).

\bibitem[{\citenamefont{Xu et~al.}(2011)\citenamefont{Xu, He, and
  Wang}}]{Xu2011}
\bibinfo{author}{\bibfnamefont{X.-D.} \bibnamefont{Xu}},
  \bibinfo{author}{\bibfnamefont{J.-H.} \bibnamefont{He}}, \bibnamefont{and}
  \bibinfo{author}{\bibfnamefont{B.}~\bibnamefont{Wang}},
  \bibinfo{journal}{Phys.Lett. B} \textbf{\bibinfo{volume}{701}},
  \bibinfo{pages}{513} (\bibinfo{year}{2011}).

\bibitem[{\citenamefont{Salvatelli et~al.}(2013)\citenamefont{Salvatelli,
  Marchini, Lopez-Honorez, and Mena}}]{Salvatelli2013}
\bibinfo{author}{\bibfnamefont{V.}~\bibnamefont{Salvatelli}},
  \bibinfo{author}{\bibfnamefont{A.}~\bibnamefont{Marchini}},
  \bibinfo{author}{\bibfnamefont{L.}~\bibnamefont{Lopez-Honorez}},
  \bibnamefont{and} \bibinfo{author}{\bibfnamefont{O.}~\bibnamefont{Mena}}
  (\bibinfo{year}{2013}).

\bibitem[{\citenamefont{Olivares
  et~al.}(2008{\natexlab{b}})\citenamefont{Olivares, Atrio-Barandela, and
  Pav\'on}}]{Olivares2008b}
\bibinfo{author}{\bibfnamefont{G.}~\bibnamefont{Olivares}},
  \bibinfo{author}{\bibfnamefont{F.}~\bibnamefont{Atrio-Barandela}},
  \bibnamefont{and} \bibinfo{author}{\bibfnamefont{D.}~\bibnamefont{Pav\'on}},
  \bibinfo{journal}{Phys. Rev. D} \textbf{\bibinfo{volume}{77}},
  \bibinfo{pages}{103520} (\bibinfo{year}{2008}{\natexlab{b}}).

\bibitem[{Pla(2013{\natexlab{a}})}]{PlanckCollaboration2013a}
\bibinfo{journal}{Planck intermediate results. XIII.}
  (\bibinfo{year}{2013}{\natexlab{a}}), \eprint{arXiv:1303.5090}.

\bibitem[{\citenamefont{Kashlinsky et~al.}(2008)\citenamefont{Kashlinsky,
  Atrio-Barandela, Kocevski, and Ebeling}}]{Kashlinsky2008}
\bibinfo{author}{\bibfnamefont{A.}~\bibnamefont{Kashlinsky}},
  \bibinfo{author}{\bibfnamefont{F.}~\bibnamefont{Atrio-Barandela}},
  \bibinfo{author}{\bibfnamefont{D.}~\bibnamefont{Kocevski}}, \bibnamefont{and}
  \bibinfo{author}{\bibfnamefont{H.}~\bibnamefont{Ebeling}},
  \bibinfo{journal}{Astrophys.J.} \textbf{\bibinfo{volume}{686}},
  \bibinfo{pages}{L49} (\bibinfo{year}{2008}).

\bibitem[{\citenamefont{Kashlinsky et~al.}(2010)\citenamefont{Kashlinsky,
  Atrio-Barandela, Ebeling, Edge, and Kocevski}}]{Kashlinsky2010}
\bibinfo{author}{\bibfnamefont{A.}~\bibnamefont{Kashlinsky}},
  \bibinfo{author}{\bibfnamefont{F.}~\bibnamefont{Atrio-Barandela}},
  \bibinfo{author}{\bibfnamefont{H.}~\bibnamefont{Ebeling}},
  \bibinfo{author}{\bibfnamefont{A.}~\bibnamefont{Edge}}, \bibnamefont{and}
  \bibinfo{author}{\bibfnamefont{D.}~\bibnamefont{Kocevski}},
  \bibinfo{journal}{Astrophys.J.} \textbf{\bibinfo{volume}{712}},
  \bibinfo{pages}{L81} (\bibinfo{year}{2010}).

\bibitem[{\citenamefont{Kashlinsky et~al.}(2011)\citenamefont{Kashlinsky,
  Atrio-Barandela, and Ebeling}}]{Kashlinsky2011}
\bibinfo{author}{\bibfnamefont{A.}~\bibnamefont{Kashlinsky}},
  \bibinfo{author}{\bibfnamefont{F.}~\bibnamefont{Atrio-Barandela}},
  \bibnamefont{and} \bibinfo{author}{\bibfnamefont{H.}~\bibnamefont{Ebeling}},
  \bibinfo{journal}{Astrophys.J.} \textbf{\bibinfo{volume}{732}},
  \bibinfo{pages}{1} (\bibinfo{year}{2011}).

\bibitem[{\citenamefont{Hand et~al.}(2012)\citenamefont{Hand, Addison, Aubourg,
  Battaglia, Battistelli, Bizyaev, Bond, Brewington, Brinkmann, Brown
  et~al.}}]{Hand2012}
\bibinfo{author}{\bibfnamefont{N.}~\bibnamefont{Hand}},
  \bibinfo{author}{\bibfnamefont{G.~E.} \bibnamefont{Addison}},
  \bibinfo{author}{\bibfnamefont{E.}~\bibnamefont{Aubourg}},
  \bibinfo{author}{\bibfnamefont{N.}~\bibnamefont{Battaglia}},
  \bibinfo{author}{\bibfnamefont{E.~S.} \bibnamefont{Battistelli}},
  \bibinfo{author}{\bibfnamefont{D.}~\bibnamefont{Bizyaev}},
  \bibinfo{author}{\bibfnamefont{J.~R.} \bibnamefont{Bond}},
  \bibinfo{author}{\bibfnamefont{H.}~\bibnamefont{Brewington}},
  \bibinfo{author}{\bibfnamefont{J.}~\bibnamefont{Brinkmann}},
  \bibinfo{author}{\bibfnamefont{B.~R.} \bibnamefont{Brown}},
  \bibnamefont{et~al.}, \bibinfo{journal}{Phys. Rev. Lett.}
  \textbf{\bibinfo{volume}{109}}, \bibinfo{pages}{041101}
  (\bibinfo{year}{2012}).

\bibitem[{\citenamefont{Mroczkowski et~al.}(2012)\citenamefont{Mroczkowski,
  Dicker, Sayers, Reese, Mason, Czakon, Romero, Young, Devlin, Golwala
  et~al.}}]{Mroczkowski2012}
\bibinfo{author}{\bibfnamefont{T.}~\bibnamefont{Mroczkowski}},
  \bibinfo{author}{\bibfnamefont{S.}~\bibnamefont{Dicker}},
  \bibinfo{author}{\bibfnamefont{J.}~\bibnamefont{Sayers}},
  \bibinfo{author}{\bibfnamefont{E.~D.} \bibnamefont{Reese}},
  \bibinfo{author}{\bibfnamefont{B.}~\bibnamefont{Mason}},
  \bibinfo{author}{\bibfnamefont{N.}~\bibnamefont{Czakon}},
  \bibinfo{author}{\bibfnamefont{C.}~\bibnamefont{Romero}},
  \bibinfo{author}{\bibfnamefont{A.}~\bibnamefont{Young}},
  \bibinfo{author}{\bibfnamefont{M.}~\bibnamefont{Devlin}},
  \bibinfo{author}{\bibfnamefont{S.}~\bibnamefont{Golwala}},
  \bibnamefont{et~al.}, \bibinfo{journal}{ApJ} \textbf{\bibinfo{volume}{761}},
  \bibinfo{pages}{47} (\bibinfo{year}{2012}).

\bibitem[{\citenamefont{Bull et~al.}(2011)\citenamefont{Bull, Clifton, and
  Ferreira}}]{Bull2011}
\bibinfo{author}{\bibfnamefont{P.}~\bibnamefont{Bull}},
  \bibinfo{author}{\bibfnamefont{T.}~\bibnamefont{Clifton}}, \bibnamefont{and}
  \bibinfo{author}{\bibfnamefont{P.~G.} \bibnamefont{Ferreira}},
  \bibinfo{journal}{Phys. Rev. D} \textbf{\bibinfo{volume}{85}},
  \bibinfo{pages}{024002} (\bibinfo{year}{2011}).

\bibitem[{\citenamefont{Genova-Santos et~al.}(2009)\citenamefont{Genova-Santos,
  Atrio-Barandela, Muecket, and Klar}}]{Genova-Santos2009}
\bibinfo{author}{\bibfnamefont{R.}~\bibnamefont{Genova-Santos}},
  \bibinfo{author}{\bibfnamefont{F.}~\bibnamefont{Atrio-Barandela}},
  \bibinfo{author}{\bibfnamefont{J.}~\bibnamefont{Muecket}}, \bibnamefont{and}
  \bibinfo{author}{\bibfnamefont{J.}~\bibnamefont{Klar}},
  \bibinfo{journal}{Astrophys.J.} \textbf{\bibinfo{volume}{700}},
  \bibinfo{pages}{447} (\bibinfo{year}{2009}).

\bibitem[{\citenamefont{Zhang}(2010)}]{Zhang2010}
\bibinfo{author}{\bibfnamefont{P.}~\bibnamefont{Zhang}},
  \bibinfo{journal}{MNRAS Letters,} \textbf{\bibinfo{volume}{407}},
  \bibinfo{pages}{L36} (\bibinfo{year}{2010}).

\bibitem[{\citenamefont{Kodama and Sasaki}(1984)}]{Kodama1984}
\bibinfo{author}{\bibfnamefont{H.}~\bibnamefont{Kodama}} \bibnamefont{and}
  \bibinfo{author}{\bibfnamefont{M.}~\bibnamefont{Sasaki}},
  \bibinfo{journal}{Progress of Theoretical Physics Supplement}
  \textbf{\bibinfo{volume}{78}}, \bibinfo{pages}{1} (\bibinfo{year}{1984}).

\bibitem[{\citenamefont{Ma and Bertschinger}(1995)}]{Ma1995}
\bibinfo{author}{\bibfnamefont{C.-P.} \bibnamefont{Ma}} \bibnamefont{and}
  \bibinfo{author}{\bibfnamefont{E.}~\bibnamefont{Bertschinger}},
  \bibinfo{journal}{Astrophys.J.} \textbf{\bibinfo{volume}{455}},
  \bibinfo{pages}{7} (\bibinfo{year}{1995}).

\bibitem[{Pla(2013{\natexlab{b}})}]{PlanckCollaboration2013f}
\bibinfo{journal}{Planck 2013 results. XV.}
  (\bibinfo{year}{2013}{\natexlab{b}}), \eprint{arXiv:1303.5075}.

\bibitem[{\citenamefont{Sachs and Wolfe}(1967)}]{Sachs1967}
\bibinfo{author}{\bibfnamefont{R.~K.} \bibnamefont{Sachs}} \bibnamefont{and}
  \bibinfo{author}{\bibfnamefont{A.~M.} \bibnamefont{Wolfe}},
  \bibinfo{journal}{Astrophys. J.} \textbf{\bibinfo{volume}{147}},
  \bibinfo{pages}{73} (\bibinfo{year}{1967}).

\bibitem[{\citenamefont{Hu and Sugiyama}(1995)}]{Hu1995}
\bibinfo{author}{\bibfnamefont{W.}~\bibnamefont{Hu}} \bibnamefont{and}
  \bibinfo{author}{\bibfnamefont{N.}~\bibnamefont{Sugiyama}},
  \bibinfo{journal}{Phys. Rev. D} \textbf{\bibinfo{volume}{51}},
  \bibinfo{pages}{2599} (\bibinfo{year}{1995}).

\bibitem[{\citenamefont{Crittenden and Turok}(1996)}]{Crittenden1996}
\bibinfo{author}{\bibfnamefont{R.~G.} \bibnamefont{Crittenden}}
  \bibnamefont{and} \bibinfo{author}{\bibfnamefont{N.}~\bibnamefont{Turok}},
  \bibinfo{journal}{Phys. Rev. Lett.} \textbf{\bibinfo{volume}{76}},
  \bibinfo{pages}{575} (\bibinfo{year}{1996}).

\bibitem[{\citenamefont{Doran}(2005)}]{Doran2005}
\bibinfo{author}{\bibfnamefont{M.}~\bibnamefont{Doran}},
  \bibinfo{journal}{JCAP} \textbf{\bibinfo{volume}{0506}}, \bibinfo{pages}{011}
  (\bibinfo{year}{2005}).

\bibitem[{\citenamefont{Kashlinsky and Atrio-Barandela}(2000)}]{Kashlinsky2000}
\bibinfo{author}{\bibfnamefont{A.}~\bibnamefont{Kashlinsky}} \bibnamefont{and}
  \bibinfo{author}{\bibfnamefont{F.}~\bibnamefont{Atrio-Barandela}},
  \bibinfo{journal}{ApJ} \textbf{\bibinfo{volume}{536}}, \bibinfo{pages}{L67}
  (\bibinfo{year}{2000}).

\bibitem[{\citenamefont{Sunyaev and Zeldovich}(1972)}]{Sunyaev1972}
\bibinfo{author}{\bibfnamefont{R.~A.} \bibnamefont{Sunyaev}} \bibnamefont{and}
  \bibinfo{author}{\bibfnamefont{Y.~B.} \bibnamefont{Zeldovich}},
  \bibinfo{journal}{Comments on Astrophysics and Space Physics}
  \textbf{\bibinfo{volume}{4}}, \bibinfo{pages}{173} (\bibinfo{year}{1972}).

\bibitem[{\citenamefont{Vishniac}(1987)}]{Vishniac1987}
\bibinfo{author}{\bibfnamefont{E.~T.} \bibnamefont{Vishniac}},
  \bibinfo{journal}{Astrophys.J.} \textbf{\bibinfo{volume}{322}},
  \bibinfo{pages}{597} (\bibinfo{year}{1987}).

\bibitem[{\citenamefont{Zhang et~al.}(2004)\citenamefont{Zhang, Pen, and
  Trac}}]{Zhang2004}
\bibinfo{author}{\bibfnamefont{P.}~\bibnamefont{Zhang}},
  \bibinfo{author}{\bibfnamefont{U.-L.} \bibnamefont{Pen}}, \bibnamefont{and}
  \bibinfo{author}{\bibfnamefont{H.}~\bibnamefont{Trac}},
  \bibinfo{journal}{Mon.Not.Roy.Astron.Soc.} \textbf{\bibinfo{volume}{347}},
  \bibinfo{pages}{1224} (\bibinfo{year}{2004}).

\bibitem[{\citenamefont{Fang et~al.}(1993)\citenamefont{Fang, Bi, Xiang, and
  Boerner}}]{Fang1993}
\bibinfo{author}{\bibfnamefont{L.-Z.} \bibnamefont{Fang}},
  \bibinfo{author}{\bibfnamefont{H.}~\bibnamefont{Bi}},
  \bibinfo{author}{\bibfnamefont{S.}~\bibnamefont{Xiang}}, \bibnamefont{and}
  \bibinfo{author}{\bibfnamefont{G.}~\bibnamefont{Boerner}},
  \bibinfo{journal}{ApJ} \textbf{\bibinfo{volume}{413}}, \bibinfo{pages}{477}
  (\bibinfo{year}{1993}).

\bibitem[{\citenamefont{Hu}(2000)}]{Hu2000}
\bibinfo{author}{\bibfnamefont{W.}~\bibnamefont{Hu}},
  \bibinfo{journal}{Astrophys.J.} \textbf{\bibinfo{volume}{529}},
  \bibinfo{pages}{12} (\bibinfo{year}{2000}).

\bibitem[{\citenamefont{Ma and Fry}(2002)}]{Ma2002}
\bibinfo{author}{\bibfnamefont{C.-P.} \bibnamefont{Ma}} \bibnamefont{and}
  \bibinfo{author}{\bibfnamefont{J.~N.} \bibnamefont{Fry}},
  \bibinfo{journal}{Phys.Rev.Lett.} \textbf{\bibinfo{volume}{88}},
  \bibinfo{pages}{211301} (\bibinfo{year}{2002}).

\bibitem[{\citenamefont{Shaw et~al.}(2011)\citenamefont{Shaw, Rudd, and
  Nagai}}]{Shaw2011}
\bibinfo{author}{\bibfnamefont{L.~D.} \bibnamefont{Shaw}},
  \bibinfo{author}{\bibfnamefont{D.~H.} \bibnamefont{Rudd}}, \bibnamefont{and}
  \bibinfo{author}{\bibfnamefont{D.}~\bibnamefont{Nagai}}
  (\bibinfo{year}{2011}).

\bibitem[{\citenamefont{Coles and Jones}(1991)}]{Coles1991}
\bibinfo{author}{\bibfnamefont{P.}~\bibnamefont{Coles}} \bibnamefont{and}
  \bibinfo{author}{\bibfnamefont{B.}~\bibnamefont{Jones}},
  \bibinfo{journal}{MNRAS} \textbf{\bibinfo{volume}{248}}, \bibinfo{pages}{1}
  (\bibinfo{year}{1991}).

\bibitem[{\citenamefont{Smith et~al.}(2003)\citenamefont{Smith, Peacock,
  Jenkins, White, Frenk, Pearce, Thomas, Efstathiou, Couchmann, and
  Consortium}}]{Smith2003}
\bibinfo{author}{\bibfnamefont{R.~E.} \bibnamefont{Smith}},
  \bibinfo{author}{\bibfnamefont{J.~A.} \bibnamefont{Peacock}},
  \bibinfo{author}{\bibfnamefont{A.}~\bibnamefont{Jenkins}},
  \bibinfo{author}{\bibfnamefont{S.~D.~M.} \bibnamefont{White}},
  \bibinfo{author}{\bibfnamefont{C.~S.} \bibnamefont{Frenk}},
  \bibinfo{author}{\bibfnamefont{F.~R.} \bibnamefont{Pearce}},
  \bibinfo{author}{\bibfnamefont{P.~A.} \bibnamefont{Thomas}},
  \bibinfo{author}{\bibfnamefont{G.}~\bibnamefont{Efstathiou}},
  \bibinfo{author}{\bibfnamefont{H.~M.~P.} \bibnamefont{Couchmann}},
  \bibnamefont{and} \bibinfo{author}{\bibfnamefont{T.~V.}
  \bibnamefont{Consortium}}, \bibinfo{journal}{Mon.Not.Roy.Astron.Soc.}
  \textbf{\bibinfo{volume}{341}}, \bibinfo{pages}{1311} (\bibinfo{year}{2003}).

\bibitem[{\citenamefont{Takahashi et~al.}(2012)\citenamefont{Takahashi, Sato,
  Nishimichi, Taruya, and Oguri}}]{Takahashi2012}
\bibinfo{author}{\bibfnamefont{R.}~\bibnamefont{Takahashi}},
  \bibinfo{author}{\bibfnamefont{M.}~\bibnamefont{Sato}},
  \bibinfo{author}{\bibfnamefont{T.}~\bibnamefont{Nishimichi}},
  \bibinfo{author}{\bibfnamefont{A.}~\bibnamefont{Taruya}}, \bibnamefont{and}
  \bibinfo{author}{\bibfnamefont{M.}~\bibnamefont{Oguri}},
  \bibinfo{journal}{The Astrophysical Journal} \textbf{\bibinfo{volume}{761}},
  \bibinfo{pages}{152} (\bibinfo{year}{2012}).

\bibitem[{\citenamefont{Sievers et~al.}(2013)\citenamefont{Sievers, Hlozek,
  Nolta, Acquaviva, Addison, Ade, Aguirre, Amiri, Appel, Barrientos
  et~al.}}]{Sievers2013}
\bibinfo{author}{\bibfnamefont{J.~L.} \bibnamefont{Sievers}},
  \bibinfo{author}{\bibfnamefont{R.~A.} \bibnamefont{Hlozek}},
  \bibinfo{author}{\bibfnamefont{M.~R.} \bibnamefont{Nolta}},
  \bibinfo{author}{\bibfnamefont{V.}~\bibnamefont{Acquaviva}},
  \bibinfo{author}{\bibfnamefont{G.~E.} \bibnamefont{Addison}},
  \bibinfo{author}{\bibfnamefont{P.~A.~R.} \bibnamefont{Ade}},
  \bibinfo{author}{\bibfnamefont{P.}~\bibnamefont{Aguirre}},
  \bibinfo{author}{\bibfnamefont{M.}~\bibnamefont{Amiri}},
  \bibinfo{author}{\bibfnamefont{J.~W.} \bibnamefont{Appel}},
  \bibinfo{author}{\bibfnamefont{L.~F.} \bibnamefont{Barrientos}},
  \bibnamefont{et~al.} (\bibinfo{year}{2013}).

\bibitem[{\citenamefont{Reichardt et~al.}(2012)\citenamefont{Reichardt, Shaw,
  Zahn, Aird, Benson, Bleem, Carlstrom, Chang, Cho, Crawford
  et~al.}}]{Reichardt2011}
\bibinfo{author}{\bibfnamefont{C.~L.} \bibnamefont{Reichardt}},
  \bibinfo{author}{\bibfnamefont{L.}~\bibnamefont{Shaw}},
  \bibinfo{author}{\bibfnamefont{O.}~\bibnamefont{Zahn}},
  \bibinfo{author}{\bibfnamefont{K.~A.} \bibnamefont{Aird}},
  \bibinfo{author}{\bibfnamefont{B.~A.} \bibnamefont{Benson}},
  \bibinfo{author}{\bibfnamefont{L.~E.} \bibnamefont{Bleem}},
  \bibinfo{author}{\bibfnamefont{J.~E.} \bibnamefont{Carlstrom}},
  \bibinfo{author}{\bibfnamefont{C.~L.} \bibnamefont{Chang}},
  \bibinfo{author}{\bibfnamefont{H.~M.} \bibnamefont{Cho}},
  \bibinfo{author}{\bibfnamefont{T.~M.} \bibnamefont{Crawford}},
  \bibnamefont{et~al.}, \bibinfo{journal}{ApJ} \textbf{\bibinfo{volume}{755}},
  \bibinfo{pages}{70} (\bibinfo{year}{2012}).

\end{thebibliography}

\end{document}